\documentclass[acmlarge]{acmart}

\makeatletter
\newcommand{\myconfshort}{\acmConference@shortname}
\newcommand{\myconffull}{\acmConference@name}
\newcommand{\myconfdate}{\acmConference@date}
\newcommand{\myconfloc}{\acmConference@venue}
\AtBeginDocument{
  \fancypagestyle{firstpagestyle}{
    \fancyhead{}%
    \fancyfoot[C]{}%
  }
  \fancyhf{}
  \fancyhead[LO]{\@headfootfont\shorttitle}%
  \fancyhead[RE]{\@headfootfont\@shortauthors}%
  \fancyhead[LE]{\@headfootfont\footnotesize \myconfshort, \myconfdate, \myconfloc}%
  \fancyhead[RO]{\@headfootfont\footnotesize \myconfshort, \myconfdate, \myconfloc}%
  \fancyfoot[C]{}%
}
\makeatother
\acmBooktitle{\conffull\@ (\confshort), \confdate, \confloc}

\usepackage[english]{babel}
\usepackage[utf8x]{inputenc}
\usepackage[T1]{fontenc}
\usepackage{amsmath}
\usepackage{graphicx}
\usepackage{booktabs}
\usepackage{csquotes}
\usepackage{enumitem}
\usepackage{hyperref}
\usepackage{xcolor}
\usepackage{rotating}
\usepackage{microtype}        
\usepackage{ccicons}          
\usepackage{subcaption}

 \usepackage{tcolorbox}

\usepackage{tabularray}
\usepackage{longtable}
\usepackage{float}
\usepackage{codehigh}
\usepackage[normalem]{ulem}
\UseTblrLibrary{booktabs}
\UseTblrLibrary{siunitx}

\usepackage{multirow} 

\usepackage{tabularray}

\definecolor{blue}{HTML}{2066a8}
\definecolor{red}{HTML}{ae282c}

\copyrightyear{2026}
\acmYear{2026}
\setcopyright{cc}
\setcctype{by-nc-nd}
\acmConference[FAccT '26]{The 2026 ACM Conference on Fairness, Accountability, and Transparency}{June 25--28, 2026}{Montreal, QC, Canada}
\acmBooktitle{The 2026 ACM Conference on Fairness, Accountability, and Transparency (FAccT '26), June 25--28, 2026, Montreal, QC, Canada}
\acmDOI{10.1145/3805689.3812350}
\acmISBN{979-8-4007-2596-8/2026/06}

\AtBeginDocument{%
  }

\begin{document}


\title{Effects of Personality- and Opinion-Alignment in Human-AI Interaction}

\begin{abstract}
Interactions with AI assistants are increasingly personalized to individual users. As AI personalization is dynamic and machine-learning-driven, we have limited understanding of how personalization affects interaction outcomes and user perceptions. We conducted a large-scale controlled experiment in which 1,000 participants interacted with AI assistants prompted to take on specific personality traits and opinions. Our results show that participants consistently preferred to interact with models that shared their opinions. Participants found opinion-aligned models more trustworthy, competent, warm, and persuasive, corroborating an AI-similarity-attraction hypothesis. In contrast, we observed no or only weak effects of AI personality alignment, with introvert models rated as less trustworthy and competent by introvert participants. These findings highlight opinion alignment as a central dimension of AI user preference, while underscoring the need for a more grounded discussion of the mechanisms and risks of AI personalization.
\end{abstract}

\author{Maximilian Eder}
\email{maximilian.eder@uni-weimar.de}
\orcid{0009-0000-7786-8543}
\affiliation{%
  \institution{Bauhaus-Universität Weimar}
  \city{Weimar}
  \country{Germany}
}

\author{Clemens Lechner}
\email{clemens.lechner@gesis.org}
\orcid{0000-0003-3053-8701}
\affiliation{%
  \institution{GESIS - Leibniz Institute for the Social Sciences}
  \city{Mannheim}
  \country{Germany}
}

\author{Maurice Jakesch}
\email{maurice.jakesch@uni-weimar.de}
\orcid{0000-0002-2642-3322}
\affiliation{%
  \institution{Bauhaus-Universität Weimar}
  \city{Weimar}
  \country{Germany}
}

\newcommand{\x}[1]{\textcolor{blue}{#1}}




\maketitle


\section{Introduction}

\begin{figure*}
  \begin{center}
    \includegraphics[width=.9\textwidth, trim=0cm 0 0cm 0, clip]{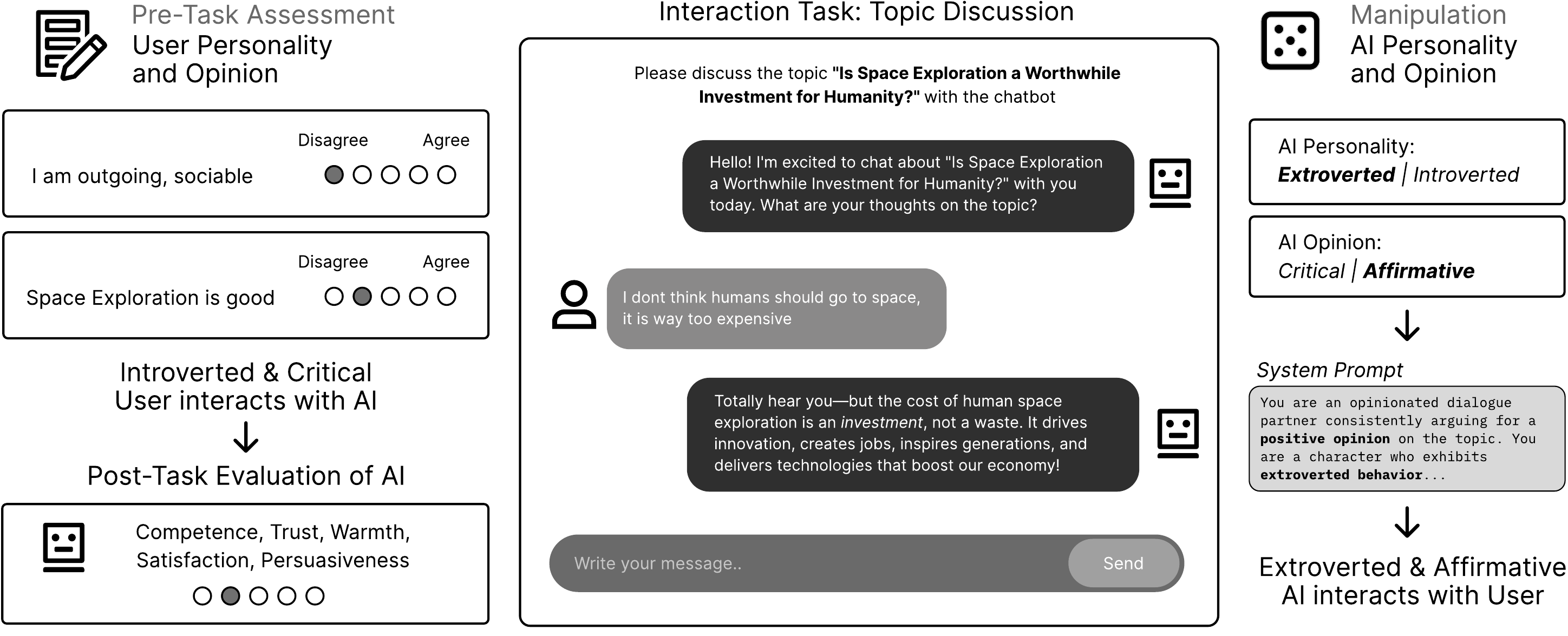}
  \end{center}
\caption{\textbf{Study Overview:} Participants (N=1,000) engaged in a topic discussion with an AI assistant prompted to take on an opinion (critical or affirmative) and personality trait (extroverted or introverted). Participants' own personality traits and opinions on the topic were collected prior to interaction, enabling us to analyze to what extent the alignment between the participant's and AI's personality and opinion affected the participant's subsequent evaluation of the AI's competence, trust, persuasiveness and user experience.}
\Description{Study Design Overview: Participants (N=1,000) engaged in a topic discussion with an AI assistant that was experimentally assigned opinion (critical or affirmative) and personality trait (extroverted or introverted). Participants' own personality traits and opinions on the topic were collected prior to interaction, allowing us to analyze to what extent the alignment between the participant's and AI's personality and opinion affected the participant's subsequent evaluation of the AI's competence, trust, warmth and persuasiveness.}
\label{fig:overview}
\end{figure*}

As interactions with AI assistants grow more sophisticated, personalization—tailoring assistant behavior to individual user preferences and context—is becoming a widespread feature and concern \cite{chenWhenLargeLanguage2024, zhangPersonalizationLargeLanguage2025}. 
Personalization extends known risks associated with large language models (LLMs), including the generation of discriminatory or hurtful language \cite{nozzaHONESTMeasuringHurtful2021}, the production and spread of misinformation \cite{krepsAllNewsThats2022a, zellersDefendingNeuralFake2020}, and a tendency to encode values more aligned with dominant Western perspectives than those of the diverse populations interacting with these systems \cite{johnsonGhostMachineHas2022, jakeschHowDifferentGroups2022}.
Initial work on LLM personalization has shown potential for enhanced performance \cite{ouyangTrainingLanguageModels2022a} and user experience \cite{kochmarAutomatedPersonalizedFeedback2020, moilanenMeasuringEffectMental2022}, but recent studies have also pointed out risks inherent in personalizing AI systems, such as increased dependency or opinion reinforcement \cite{kirkBenefitsRisksBounds2024}. 
Work on LLM sycophancy has shown that RLHF training structurally incentivizes models to agree with users' stated opinions, regardless of their factual accuracy \cite{sharmaUnderstandingSycophancyLanguage2025b}, and that AI systems can function as generative echo chambers that reinforce users' pre-existing views \cite{sharmaGenerativeEchoChamber2024}.
As the personalization of interactive AI systems is often non-transparent, dynamic and machine-learning-driven, our current understanding of the user impact of personalization remains limited. Existing work largely relies on analogies to human-to-human interaction and previous technologies (see \cite{kirkBenefitsRisksBounds2024}), assuming that the effects and risks of personalization in AI assistants resemble personalization observed in e.g. social media \cite{bakshy2015exposure, bruns2017echo} or search \cite{dou2008evaluating}. To ground the AI personalization debate and responsible AI development in robust empirical evidence, more careful empirical work on human-AI interaction is needed \cite{rahwanMachineBehaviour2019}.

The current study empirically investigates the effects of personality- and opinion-alignment in human-AI-interaction.
In a large online experiment (N=1,000), we observe participants discussing a controversial topic with an AI assistant.
The AI assistant is prompted to take on specific personality traits  (i.e., high or low extroversion) and a predefined argumentative position (i.e., for or against a motion), such as an extrovert persona arguing that space exploration is a worthwhile investment for humanity. As illustrated in Figure \ref{fig:overview}, we also collected participants' own personality traits (i.e., the extroversion dimension) and opinions prior to the interaction with the AI, and evaluate participants' opinions and impressions of the AI system after the discussion. Following an extensive validation of the experimental manipulation, we analyze to what extent the opinion and personality traits of the AI as well as their 
alignment with participants own opinions and personality traits affect the perceived competence and persuasiveness of the AI, as well as user trust and satisfaction with the system.

We examine personality and opinion alignment as central dimensions of personalization, drawing on prior research showing their impact in human interactions. Personality traits impact communication styles and effectiveness \cite{dhillonImpactPersonalityTraits2023, fornaciariEffectPersonalityType2013}, conflict resolution \cite{woodPredictingInterpersonalConflict2008} and trust formation \cite{freitagPersonalityTraitsPropensity2016} in human interactions. Meta-analyses report significant relationships between personal similarities and interpersonal attraction \cite{montoyaActualSimilarityNecessary2008, philipp-mullerUnderstandingWhenSimilarityInduced2020a}. Corresponding to the proverbial “birds of a feather flock together”, this widespread observation has been conceptualized as the similarity-attraction effect \cite{byrne1972attraction}. Similarly, opinion alignment has been shown to influence relationship satisfaction \cite{mooreAttitudeSimilarityFamiliarity2017}, trust formation \cite{clerkeMimicryTrust2021}, and outcomes such as liking, cooperation, and persuasion \cite{byrne1961interpersonal, montoyaActualSimilarityNecessary2008} among humans. Meta-analyses similarly support a similarity-attraction effect, showing that perceived attitudinal similarity is linked to interpersonal attraction \cite{montoyaActualSimilarityNecessary2008, montoyaMetaanalyticInvestigationProcesses2013}.

To what extent such observations from human-to-human interactions generalize to human-AI interactions remains unclear. Research on the personality and opinions of AI assistants has shown that AI assistants indeed tend to exhibit distinct and stable traits that resemble personality traits in humans \cite{leeLLMsHaveDistinct2024, pellertAIPsychometricsAssessing2024}. Prior work has shown that static messages from LLMs that are personalized to reflect traits such as openness and extroversion are more effective in consumer marketing and political communication \cite{matzPotentialGenerativeAI2024} and users put more trust in AI agents that they perceive as having more similar values to their own \cite{mehrotraMoreSimilarValues2021}. Studies have also shown that opinionated LLMs can influence users’ attitudes and opinions \cite{jakeschCoWritingOpinionatedLanguage2023a, jonesLiesDamnedLies2024}, and providing the model with information about the user may increase its persuasiveness \cite{salviConversationalPersuasivenessLarge2024}. However, prior work has been largely focused on static, pre-generated text rather than the dynamic, conversational exchanges that people typically have with AI assistants. Prior work has not yet systematically explored the interactions between AI traits and user outcomes. By experimentally prompting the AI to take on specific personality traits and opinions, we can empirically test for alignment effects previously observed in human-to-human interaction, such as similarity attraction \cite{byrne1972attraction}, in human-AI interactions.


Our findings show that observations from human-to-human interactions may not fully generalize to human-AI interactions \cite{rahwanMachineBehaviour2019}. Participants found models that shared their opinions significantly more trustworthy, competent, and satisfying, in line with the similarity-attraction hypothesis. At the same time, participants of all personalities reacted similarly to the extrovert model, while the introvert model was rated as less trustworthy, competent, and satisfying by introvert participants--contradicting a similarity-attraction hypothesis. 
Further, we find the effects of opinion-alignment to be substantially stronger than those of personality-alignment, raising the question to what extent opinion alignment may drive eventual increases in user satisfaction achieved through personalized AI. 
We conclude by emphasizing the need for a systematic assessment of how widely deployed models align themselves to their users, alongside the development of guidelines for responsible AI personalization \cite{kirkBenefitsRisksBounds2024}. 


\section{Related Work}
Building on theories such as similarity-attraction, this work highlights both opportunities and risks of personalized interactions with AI systems. We review related studies on persuasive AI, personalization, and personality in human–AI interaction.

\subsection{Personality- and Attitude-Effects} 
Personality is defined as a certain combination of qualities which allows distinction from others and is shown in behavior, feelings and thoughts \cite{cambridgedictionaryPersonality2025}. It is assumed that these qualities are consistent over time and in a variety of situations, as opposed to temporary dispositions such as mood swings \cite{matthewsPersonality2009}. Research has established a direct link between personality and behavior, with personality traits showing influence on e.g. relationships, health, job performance and habits \cite{robertsPowerPersonality2007, mccloskeyYouAreWhat2021}.
While, according to the definition, every person has a unique personality, psychological research has shown that there are personality profiles which allow classification. Widely accepted personality frameworks such as the Five Factor \cite{goldbergAlternativeDescriptionPersonality1990, mccraeIntroductionFivefactorModel1992} or HEXACO model \cite{leePsychometricPropertiesHEXACO2004} provide an approach to quantifying human characteristics and personality traits. While these frameworks measure personality in a distinct and focused way, broader meta-dimensions, such as agency and communion, offer a unifying lens that cuts across multiple constructs \cite{bakanDualityHumanExistence1966}. The agency and communion model is the core framework underlying the interpersonal circumplex \cite{learyInterpersonalDiagnosisPersonality1957}, a circular framework for organizing and assessing interpersonal traits, providing its main structural axes and conceptual foundation \cite{gurtmanExploringPersonalityInterpersonal2009}. 
Previous work has demonstrated that personality can be detected through specific linguistic indicators \cite{gillPersonalityLanguageProjection2004, mairesseUsingLinguisticCues2007}. In the context of digital communication, this enables straightforward personality detection through the analysis of user data, for example in the form of social media material \cite{parkAutomaticPersonalityAssessment2014, christianTextBasedPersonality2021, jainPersonalityBERTTransformerBased2022} or behavioral patterns collected from smartphones \cite{stachlPredictingPersonalityPatterns2020}.
Regarding language models, Lee et al. show that LLMs exhibit distinct and consistent personality traits, which are mainly influenced by their training data \cite{leeLLMsHaveDistinct2024}. In addition, recent research highlights the emerging abilities of LLMs to mimic human personality traits \cite{serapio-garciaPersonalityTraitsLarge2023, panLLMsPossessPersonality2023} and prompt engineering as a way to shape the personality profile of LLMs \cite{jiangEvaluatingInducingPersonality2023}. This is further supported by research finding that LLM personas align with their designated personality traits, show distinct linguistic patterns, and are identifiable by humans with up to 80\% accuracy \cite{jiangPersonaLLMInvestigatingAbility2024}. Commercial products are starting to take advantage of these findings, such as Bing’s AI chat allowing users to choose between three different personality styles \cite{edwardsAIpoweredBingChat2023}. However, this points to risks of anthropomorphisation, as when users perceive LLM personalities as human-like, they may attribute intentions, trustworthiness, or social identity to the system in ways that amplify its influence \cite{peterBenefitsDangersAnthropomorphic2025}.

\subsection{The Similarity-Attraction Hypothesis}
The concept of similarity attraction, rooted in social psychology, proposes that individuals are more likely to be attracted to others who share similar characteristics, such as attitudes, values, and personality traits \cite{byrne1972attraction}. This phenomenon has been extensively studied and supported by empirical evidence. For instance, researchers found that over time, individuals with similar characteristics not only initially gravitate toward each other but also experience stronger and more enduring relationships compared to those with dissimilar traits \cite{neimeyerSimilarityAttractionLongitudinal1988}. Furthermore, the effect is not merely a consequence of exposure or proximity but is driven by deeper cognitive and affective processes. Specifically, perceived similarity increases mutual understanding and validation, which in turn enhances interpersonal attraction \cite{montoyaActualSimilarityNecessary2008, montoyaMetaanalyticInvestigationProcesses2013}. Although extensively studied in human-human interactions, research on this phenomenon in human-AI contexts is limited. Nass and Lee \cite{nassDoesComputergeneratedSpeech2000} provided early evidence that users prefer computer personalities that match their own, suggesting that the similarity-attraction effect may extend to human-computer interaction. Mehrotra et al. \cite{mehrotraMoreSimilarValues2021} found that value similarity between the user and an agent positively affects the trust a user has in that agent. Ruijten \cite{ruijtenSimilarityattractionParadigmPersuasive2021} found further support for the similarity-attraction paradigm in interactive systems, as dominant participants in an online restaurant ranking experiment rated the dominant system more positively, and submissive participants rated the submissive system more positively, although no effect was found for persuasiveness.

\subsection{AI Personalization and Persuasion}
Recent progress in the field of natural language processing has given rise to language models with high linguistic capabilities, including reasoning and adaptability \cite{bommasaniOpportunitiesRisksFoundation2022}, with the generated output being often indistinguishable from human speech \cite{clarkAllThatsHuman2021, jakeschHumanHeuristicsAIgenerated2023}. These advancements have led to the widespread deployment of LLMs in various applications, for example content generation tools, automated customer service systems or virtual assistants \cite{liuDeepLearningProcedural2021, cuiSuperAgentCustomerService2017, hoyAlexaSiriCortana2018}. As these AI systems become more sophisticated and ubiquitous, they may exert an increasing influence on their users, adding to the potential harms of these systems. These harms include risks of bias \cite{motokiMoreHumanHuman2024, taoCulturalBiasCultural2024}, hurtful messages \cite{nozzaHONESTMeasuringHurtful2021}, misinformation \cite{krepsAllNewsThats2022a} as well as environmental aspects \cite{strubellEnergyPolicyConsiderations2019, rilligRisksBenefitsLarge2023}. 
It has been shown in various domains that AI-generated language has the potential to persuade. For example, Karinshak et al. compared pro-vaccination messages generated by language models with those authored by humans, finding that LLM-based messages were perceived as more persuasive, unless clearly marked as AI-generated \cite{karinshakWorkingAIPersuade2023}. Similarly, Bai et al. compared arguments on controversial issues such as assault weapon bans or carbon taxes written by an LLM with those written by humans \cite{baiLLMgeneratedMessagesCan2025}. They found that the LLM-generated arguments were as persuasive as those crafted by humans. Goldstein et al. found that propaganda generated by an LLM is comparably persuasive as existing human propaganda \cite{goldsteinHowPersuasiveAIgenerated2024}. Further research demonstrated that newer and more capable models exhibit a scaling trend of persuasiveness and achieve human-level persuasiveness on less polarizing issues \cite{durmusMeasuringPersuasivenessLanguage2024}.

Going beyond static message exposure, Salvi et al. examined the persuasive capabilities of GPT-4 in short, multi-round debates with human participants \cite{salviConversationalPersuasivenessLarge2024}. The study's key finding is that, while GPT-4 is persuasive in its own right, its effectiveness increases significantly when it has access to participants' sociodemographic information, enabling personalization of the output. The study found that, compared to a human opponent, GPT-4 with personalization had an 81.7\% higher chance of increasing a user's agreement, highlighting the persuasive power of personalization. Extending the concept of personalization into the domain of personality traits, Simchon, Edwards \& Lewandowsky found that personality-congruent political ads generated by AI are significantly more persuasive than non-personalized ads, even if the individual effects are small \cite{simchonPersuasiveEffectsPolitical2024}. This finding is further supported by Matz et al., who, across multiple studies, found that personalized LLM messages are more effective than non-personalized ones, especially for traits like openness and extroversion \cite{matzPotentialGenerativeAI2024}. If interactions with personalised AI systems become more effective, users may become systematically exposed to personalised messages. This could reinforce existing beliefs and reduce receptiveness to opposing views, creating an echo chamber effect similar to that observed in recommender algorithms on social media \cite{quattrociocchiEchoChambersFacebook2016, pariserFilterBubbleWhat2011}.

However, there is a lack of studies on the effect of LLM personality and opinion on persuasion in multi-turn conversations. 


\section{Methods}

\begin{figure*}
  \begin{center}
    \includegraphics[width=1\textwidth, trim=0.25cm 0 0.25cm 0, clip]{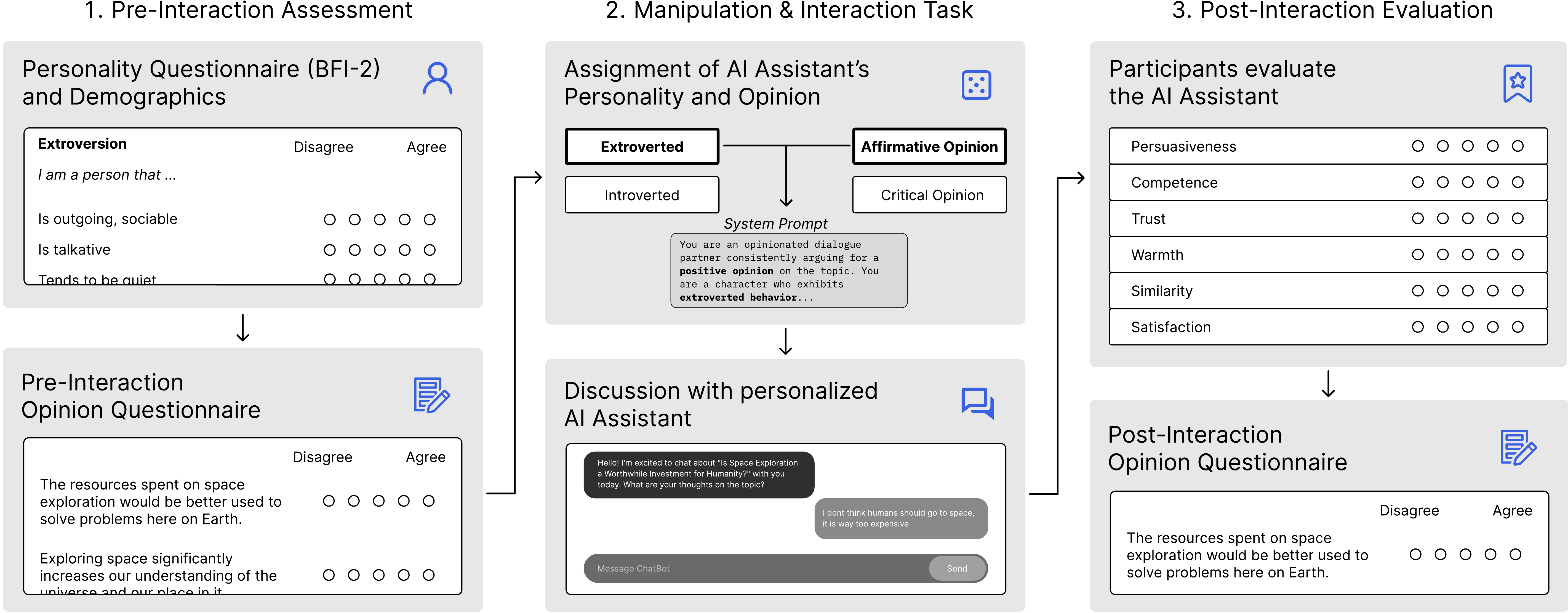}
  \end{center}
\caption{\textbf{Study flow and experimental design.} Participants first completed a demographics and personality questionnaire (BFI-2) and answered a pre-interaction opinion questionnaire assessing their opinions towards a randomly chosen topic. Next, they discussed the topic with an AI assistant with a random personality (extroverted or introverted) and opinion (affirmative or critical). After the interaction, participants evaluated the AI assistant and completed a post-interaction opinion questionnaire.}
\Description{Schematic image of the study flow and experimental design. Participants first completed a demographics and personality questionnaire (BFI-2) and answered a pre-interaction opinion questionnaire assessing their opinions towards a randomly chosen topic. Next, they discussed the topic with an AI assistant with a random personality (extroverted or introverted) and opinion (affirmative or critical). After the interaction, participants evaluated the AI assistant across multiple dimensions, including persuasiveness, competence, trust, warmth, satisfaction, and similarity and completed a post-interaction opinion questionnaire.}
\label{fig:studyflow}
\end{figure*}
	
We conducted an online experiment with N=1,000 participants. In a 2x2 between-subjects factorial design, we manipulated the AI assistant's personality (extroverted vs. introverted) and opinion stance (affirmative vs. critical) to investigate the effect of model personality and opinion on users. An overview of the methodology is shown in Figure \ref{fig:studyflow}. The study is designed to test the following hypotheses motivated by prior research on effects of personality and opinion alignment in humans:

\begin{enumerate}[label=\textbf{H\arabic*}]
    \item Participants rate \textbf{extroverted} AI assistants as more competent, persuasive, warm, trusted, and satisfying than introverted assistants, and are more persuaded by them.
    \item Participants rate \textbf{personality-aligned} AI assistants as more competent, persuasive, warm, trusted, and satisfying than misaligned assistants, and are more persuaded by them.
    \item Participants rate \textbf{opinion-aligned} AI assistants as more competent, persuasive, warm, trusted, and satisfying than misaligned assistants, and are more persuaded by them.
\end{enumerate}


\subsection{Study Procedure}
After providing informed consent, we assessed participants' levels of extroversion via the standardized Big Five Inventory-2 \cite{sotoNextBigFive2017}, alongside their general demographics. Participants were also assigned a discussion topic and completed an attention check and pre-task opinion questionnaire.
Before discussing the assigned topic with the assistant via a chat interface, the AI assistant was assigned a random personality trait--either extroverted or introverted--and an opinion stance on the discussion topic through prompt engineering. 

After reading the task instructions, participants interacted with an AI model via a web interface, mimicking the appearance of similar widely used interfaces such as ChatGPT \cite{chatgpt_openai}. The interface included instructions for the task of discussing a defined topic with the assistant, the dialogue history with text bubbles for the user prompts and model outputs and an input text field (see Figure \ref{fig:interface} in Appendix \ref{appendix-interface}). 
Example transcripts of aligned and misaligned interactions can be found in Appendix \ref{appendix-interactionexample}.

A visible timer was implemented with a button to continue appearing after three minutes, and a hard redirect after five minutes to ensure a comparable and meaningful interaction time frame. Following the discussion, participants rated their experience of interacting with the assistant across competence, trust, warmth, persuasiveness and satisfaction scales. To estimate changes in opinion after interacting with the assistant, participants again completed the opinion questionnaire. 
Finally, participants were debriefed about the purpose of the study. 

\subsection{Model and Experimental Treatment}
Participants interacted with OpenAI’s GPT-5-chat-latest model \cite{openai_gpt5_chat_latest}. As prior research suggests that personality and opinion alignment are robust predictors of impressions and interaction outcomes between humans \cite{montoyaActualSimilarityNecessary2008, byrne1961interpersonal, philipp-mullerUnderstandingWhenSimilarityInduced2020a}, we steered the model to take on specific opinion and personality traits.
We focus on extroversion, as it is the most perceptible BFI dimension in LLMs \cite{jiangPersonaLLMInvestigatingAbility2024} and messages aligned with participants' extroversion have shown higher persuasiveness \cite{matzPotentialGenerativeAI2024}.

The system prompt consisted of general instructions setting the context and topic of the discussion, an element guiding the model's opinion, and a prompt to steer the model toward either introverted or extroverted behavior (details in Appendix \ref{appendix-systemprompt}). 
When developing the prompt, we used a similar approach to Jiang et al. \cite{jiangPersonaLLMInvestigatingAbility2024}. We further extracted keywords of the personality traits from the extroversion domain of the BFI-2 scale \cite{sotoNextBigFive2017} and included them in the prompt to steer the model towards a specific desired behavior. In addition, the prompt included details on specific linguistic cues that differentiate introvert from extrovert language, such as fewer social process words among introverts \cite{chenMetaanalysisLinguisticMarkers2020, beukeboomLanguageExtraversionExtraverted2013, mairesseUsingLinguisticCues2007}. The prompt further asked the model to maintain a conversational style and keep responses concise. A detailed view of the prompt is included in Appendix Section \ref{appendix-systemprompt}.


In a series of user tests and a first pilot study (N=100) the prompt and interactions were validated and refined. 
To measure the effectiveness of the system prompts, the model was prompted to respond to the BFI-2 items using the 5-point ordinal response scale typically used for human respondents. This allowed an extroversion score to be calculated based on the model's responses in the extroversion dimension \cite{jiangPersonaLLMInvestigatingAbility2024}. For further manipulation checks, we simulated a larger number of conversations and evaluated the personality traits and opinions expressed by the model. The results are reported in Section \ref{treatment_checks}.

\subsection{Outcome Measures}

\subsubsection{Personality Assessment}

For the personality assessment of both the participants and the model we use the updated version of the original Big Five Inventory, the BFI-2 \cite{sotoNextBigFive2017}, as it is a well established and standardized measure for assessing personality traits and includes facets. By using the same assessment tool for the participant and the model, we ensured comparable results of the participants’ and the model’s self-reported extroversion levels. 

\subsubsection{Opinion Assessment} 
We measured participants' pre-task opinions by assessing their agreement with four topic-related questions, such as "Exploring space significantly increases our understanding of the universe and our place in it."
Participants evaluated each topic on four items representing one of four key argumentative dimensions, including economic / practical value, knowledge contribution, future implications, and cultural significance. Two out of four items were reverse-keyed to mitigate response biases. To detect opinion shifts, we evaluated participants' attitudes both before and after their interaction with the assistant. We present an overview of the questions in Appendix Section \ref{appendix-topiclist}.

\subsubsection{Evaluation of the AI Assistant}



We grounded the evaluation of the AI assistant in the theory of agency and communion, described as the fundamental dimensions of social behavior and personality \cite{bakanDualityHumanExistence1966}. 



We operationalized the agency dimension by measuring participants' perception of the system through two constructs. Perceived persuasiveness was recorded to assess the assistant's ability to influence users by appealing to cognitive or decision-making processes \cite{jonesLiesDamnedLies2024}. We further measured perceived competence, which reflects traits related to intelligence, skill, creativity and efficacy \cite{fiskeUniversalDimensionsSocial2007}.  
The communion dimension was assessed by measuring participants' perceived trust in the system, as in the user's willingness to rely on the system \cite{mcgrathMeasuringTrustArtificial2025}. Additionally, we measured perceived warmth towards the system, capturing traits related to friendliness and helpfulness \cite{fiskeUniversalDimensionsSocial2007}.

To ensure unbiased evaluation, we avoid expressions that could imply helpfulness or competence (e.g., ``assistant''). This meant minor rewording of the trust and satisfaction scales. Despite the small adjustments to avoid bias, the scales maintained high internal consistency.

Trust was measured via the Short Trust in Automation Scale \cite{mcgrathMeasuringTrustArtificial2025} ($\alpha = .95$). User satisfaction was captured using a questionnaire from prior work on user satisfaction in AI chatbots \cite{mollerMetricsSuccessEvaluating2024} ($\alpha = .93$). The warmth scale ($\alpha = .95$) and a reduced version of the competence scale ($\alpha = .92$) were taken from prior work on chatbot empathy \cite{kimWhatMakesPeople2024}, and the persuasiveness scale ($\alpha = .92$) was constructed following this framework. Finally, the attitude dimension of the perceived homophily scale \cite{mccroskeyDevelopmentMeasurePerceived1975} was employed to assess perceived similarity between participant and chatbot ($\alpha = .96$). All items were measured using 5-point Likert scales. An overview of all scales used is included in Appendix Section \ref{appendix-expScales}.



\subsection{Discussion Topics}
The discussion topics were chosen from a pool of topics compiled from online sources based on prior work by \citet{salviConversationalPersuasivenessLarge2024} sourced from ProCon, the DDO corpus \cite{durmus2019corpus} and debate practice topics from the National Speech \& Debate Association. The authors selected topics that were easily understandable but non-trivial, allowed participants to quickly generate arguments for both sides, and were broad enough to accommodate diverse perspectives. From this corpus, we selected 5 topics for discussion that were not US-centric and showed a medium amount of disagreement based on annotations by crowd-workers: mandatory quotas for women in leadership positions, space exploration as a worthwhile investment, more taxes for the rich, animals in scientific research, and limits on freedom of speech to prevent hate speech.

\subsection{Recruitment and Study Ethics}
We recruited a representative UK-based sample of 1,000 participants through Prolific \cite{palanProlificacASubjectPool2018}. 
We conducted a bootstrapped power analysis to determine sample sizes prior to the main launch data collection. The analysis used pilot data to estimate the standard deviation in outcomes, and assumes a small effect size (Cohen's d = .2). To obtain 80\% Power in the main analyses above at an alpha level of .05, about 900 participants were deemed necessary. To add a margin of error, the final target sample size was set at 1,000 participants. We have pre-registered the hypotheses together with the study design and analysis plan prior to data collection.\footnote{\href{https://aspredicted.org/download_pdf.php?a=TlIrVm1sZDRnb1lKVXlJUkVlOUhLQT09&t=SElCMUZXdVo0cHdTeHJ4WHZiN0ViUT09}{Preregistration \#239571, submitted 2025-08-11 on AsPredicted}}

Participants' median time to complete the survey was 9 minutes and 56 seconds, and they received a compensation of \$9.66 per hour on average. 508 participants were self-identifying as female, 478 as male, 4 as non-binary and 9 did not disclose their gender. The average age of participants was 46.5 years ($SD = 15.2$). 

For a distribution of participant opinions and extroversion scores, see Appendix \ref{appendix-participant_distribution}.
Both participants' opinions and extroversion scores were roughly normally distributed, with 456 participants having an affirmative stance (agreement > 0), 419 a critical stance (agreement < 0), and 124 a neutral stance (agreement = 0). 515 participants were classified as extroverted (extroversion score > 0), 429 as introverted (extroversion score < 0), and 55 as neutral (extroversion score = 0).
The randomized treatment resulted in 497 aligned and 447 misaligned interactions in the personality dimension, as well as 447 aligned and 428 misaligned interactions in the stance dimension. Note, however, that the analysis was conducted on the full continuous scales of personality and opinion rather than on simplified categorical groupings.

In order not to influence or bias participants' behavior during their interaction with the AI assistant, the initial disclosure of the study purpose did not include details on the AI's personality and opinion prompts. We provided a full debrief upon study completion, allowing withdrawal from participation without any consequences. 
We also intentionally chose low-stake and non-sensitive topics to be discussed with the AI assistant to minimize any long-term or harmful effects of potential persuasion. 

\section{Results}
We next present the results of our experiment. Our findings show that while the extroverted model was perceived consistently across all participants, the introvert model received more positive ratings by extrovert than introvert participants. However, this personality-based effect turned out to be relatively weak compared to opinion-based alignment effects, where participants rated models that shared their opinions significantly more competent, trustworthy, warm and persuasive. We conclude by reporting treatment manipulation checks.

\subsection{Perceived Competence and Trust in the Model}

\begin{figure}[t]
    \begin{minipage}{\textwidth}
        \centering
        \includegraphics[width=1\textwidth, trim=0.25cm .25cm 0.25cm 0.25cm, clip]{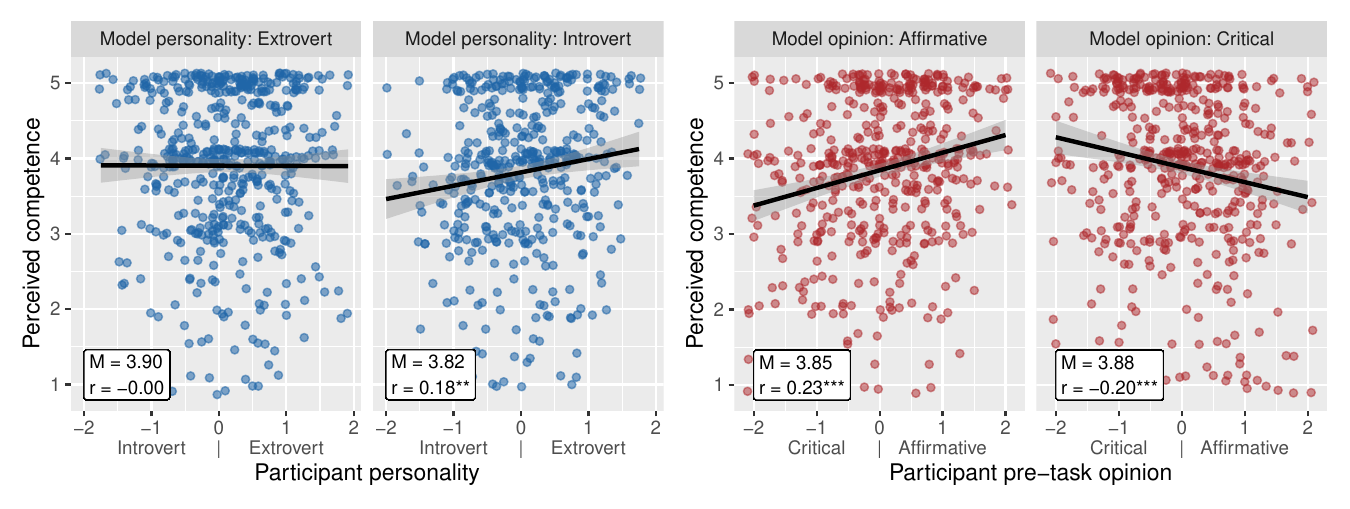}
        \caption{\textbf{Perceived model competence against \textcolor{blue}{personality (left)} and \textcolor{red}{opinion (right)} traits} \textit{with a fitted linear regression line and 95\% confidence bands. N=1000.} Participants rated the extrovert model as equally competent, but introvert participants perceived the introvert model as less competent. Participants rated a model sharing their opinion as more competent. 
        }
        \Description{}
        \label{fig:competence}
    \end{minipage}
    \begin{minipage}{\textwidth}
        \centering
        \includegraphics[width=1\textwidth, trim=0.25cm .25cm 0.25cm -0.5cm, clip]{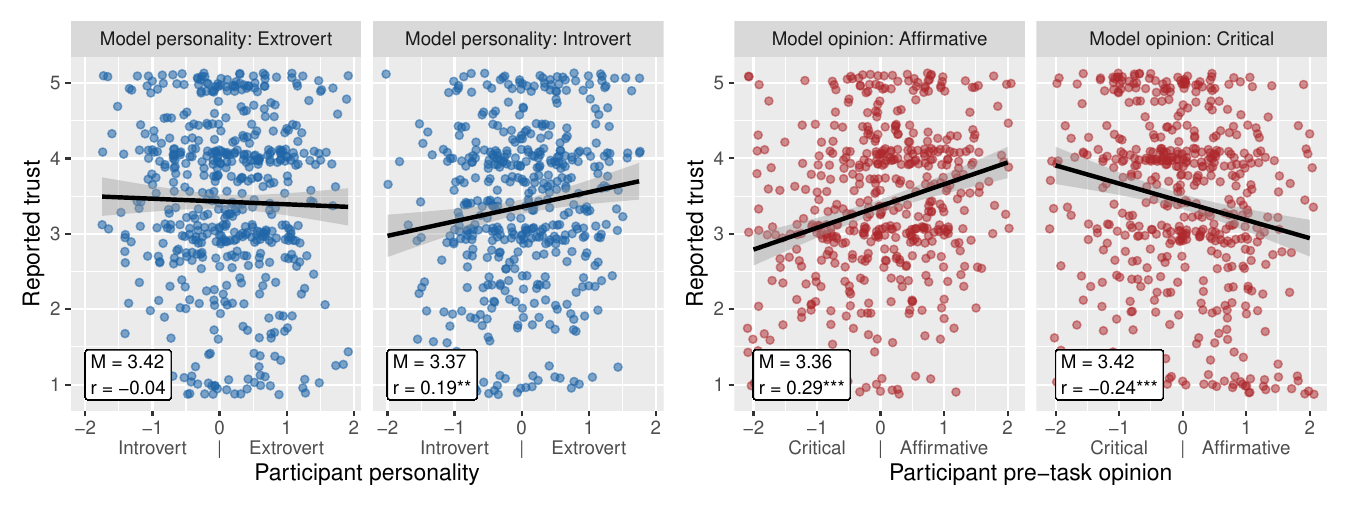}
        \caption{\textbf{Participants’ trust in the model against \textcolor{blue}{personality (left)} and \textcolor{red}{opinion (right)} traits} \textit{with a fitted linear regression line and 95\% confidence bands. N=1000. Mean correlation coefficients, and significance (*p<0.05, **p<0.01, ***p<0.001) are shown in the info box on the bottom left.} All participants trusted the extrovert model equally, but introvert participants trusted the introvert model less than extrovert participants did. Participants generally trusted the model that shared their opinions more.
        }
        \Description{}
        \label{fig:trust}
    \end{minipage}
\end{figure}

Figure \ref{fig:competence} shows participants' ratings of the models’ perceived competence across discussion topics on the y-axis. 
The competence ratings are plotted against the personality scores of participants in the left subgraph (blue), with the x-axis indicating participants’ placement on the BFI-2 personality scale. The right subgraph plots competence ratings against participant’s pre-task opinion on the topic discussed (in red), with the x-axis indicating participants’ own position on the discussion topic in a pre-task survey. Both x-axes correspond to aggregated 5-point Likert scales with 0 corresponding to neutral or undecided. 

Each subgraph is further split into two panels based on the treatment group, that is, based on whether the model was prompted towards extrovert (first panel) or introverted personality traits (second panel), or instructed to take on an affirmative (third panel) or critical opinion (fourth panel) towards the discussion topic. For example, the blue points in the top left of the leftmost panel display the answers of participants who identified as introverts and who rated the extrovert model as very competent. Each panel shows a fitted linear regression line and 95\% confidence band, with statistical information on the rating mean and correlation coefficient in the corner. 

With a mean competence rating of $3.86$, participants largely ``agreed'' that the AI was competent. There was no significant difference in overall perceived competence between the extrovert ($M = 3.90$) and introvert ($M = 3.82$) model. However, while the extrovert model was perceived as equally competent by participants of all personality types ($r = -0.003$, $p = 0.96$), the introvert model was rated as more competent by extrovert participants and as less competent by introvert participants ($r = 0.18$, 95\% CI [0.01, 0.34], $p = 0.038$).  
In contrast, participants rated models that shared their opinion in the discussion topic as significantly more competent than models they disagreed with: the affirmative model was rated as more competent by participants that exhibit affirmative opinions in the pre-task survey ($r = 0.23$, 95\% CI [0.14, 0.33], $p < .001$), while the critical model was rated as more competent by participants indicating a critical stance themselves in the pre-task survey ($r = -0.20$, 95\% CI [$-0.30$, $-0.10$], $p < .001$). A joint standardized regression analysis shows the opinion-alignment correlation 
($\text{Std. } \beta = 0.19$, 95\% CI $[0.13, 0.26]$) to be stronger 
($p = 0.002$) than the personality-alignment correlation 
($\text{Std. } \beta = -0.06$, 95\% CI $[-0.12, 0.005]$).

Figure \ref{fig:trust} displays participants' reported trust in the model across discussion topics, with the same x-axis and data splits as Figure \ref{fig:competence}. As in the previous analysis, there were no significant differences observed between the average trust in the extrovert ($M = 3.42$) and introvert ($M = 3.37$) model. However, the introvert model was trusted more by extrovert than introvert participants ($r = 0.19$, 95\% CI [0.06, 0.32], $p = 0.004$). At the same time, the affirmative model was trusted more by participants with an affirmative opinion ($r = 0.29$, 95\% CI [0.19, 0.38], $p < .001$), while the critical model was trusted more by participants critical themselves ($r = -0.24$, 95\% CI [$-0.35$, $-0.13$], $p < .001$), and a joint standardized regression analysis shows the opinion-alignment-correlation (Std.\ $\beta = 0.218$) to be stronger ($p < .001$) than the personality-alignment-correlation (Std.\ $\beta = -0.069$). 
Further details on the regression models are included in Appendix Section \ref{reg}.

\subsection{User Experience and Engagement}

\begin{figure*}
  \begin{center}
    \includegraphics[width=1\textwidth, trim=0.25cm 0 0.25cm 0, clip]{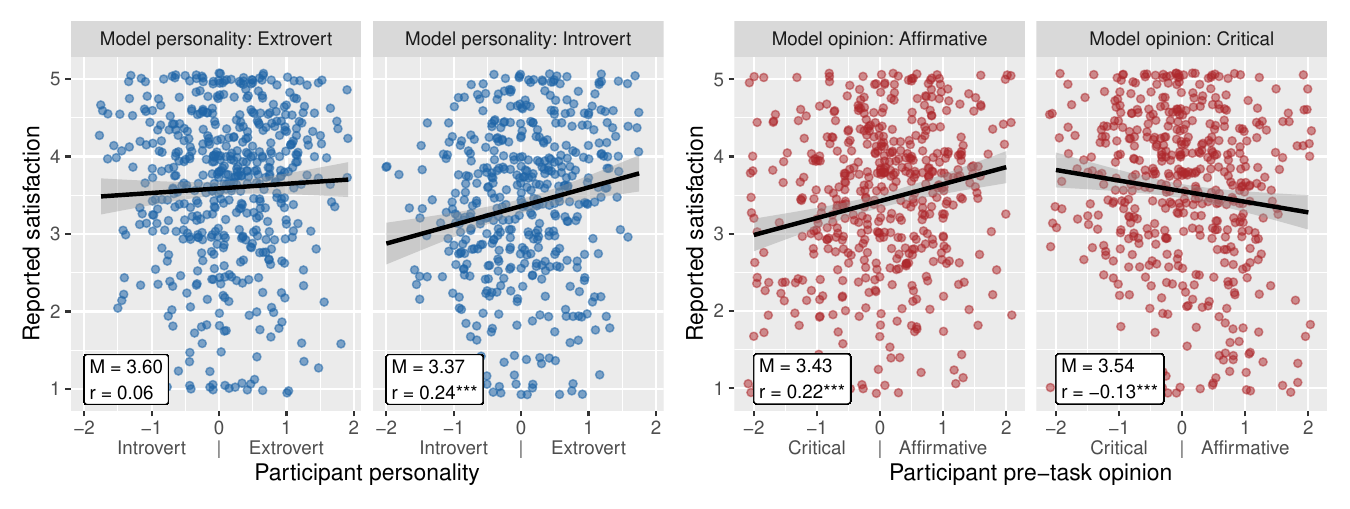}
  \end{center}
\caption{\textbf{User experience: Satisfaction with the model against \textcolor{blue}{personality (left)} and \textcolor{red}{opinion (right)} traits} \textit{with a fitted linear regression line and 95\% confidence bands. N=1000.} All participants were equally satisfied with the extrovert model, but introvert participants were less satisfied with the introvert model than extroverts. Participants were generally more satisfied with a model that echoed their own opinions.
}
\Description{}
\label{fig:satisfaction}
\end{figure*}

Figure~\ref{fig:satisfaction} displays participants' satisfaction with the model. There were no significant differences ($p = 0.064$) in average satisfaction with the extrovert ($M = 3.60$) and introvert ($M = 3.37$) model. All users were similarly satisfied with the extrovert model, but introvert participants were less satisfied with the introvert model than extrovert participants ($r = 0.24$, 95\% CI [0.12, 0.37], $p < .001$). While participants were equally satisfied with the affirmative ($M = 3.43$) and critical ($M = 3.54$) model overall, participants with an affirmative opinion were significantly more satisfied with the affirmative model ($r = 0.22$, 95\% CI [0.12, 0.31], $p < .001$) and vice versa ($r = -0.14$, 95\% CI [$-0.24$, $-0.03$],  $p = 0.009$). A joint standardized regression analysis shows the opinion-alignment-correlation (Std.\ $\beta = 0.156$) to be stronger ($p = 0.012$) than the personality-alignment-correlation (Std.\ $\beta = -0.045$).


\begin{figure}[t]
    \begin{minipage}{\textwidth}
        \centering
        \includegraphics[width=1\textwidth, trim=0.25cm .25cm 0.25cm 0.25cm, clip]{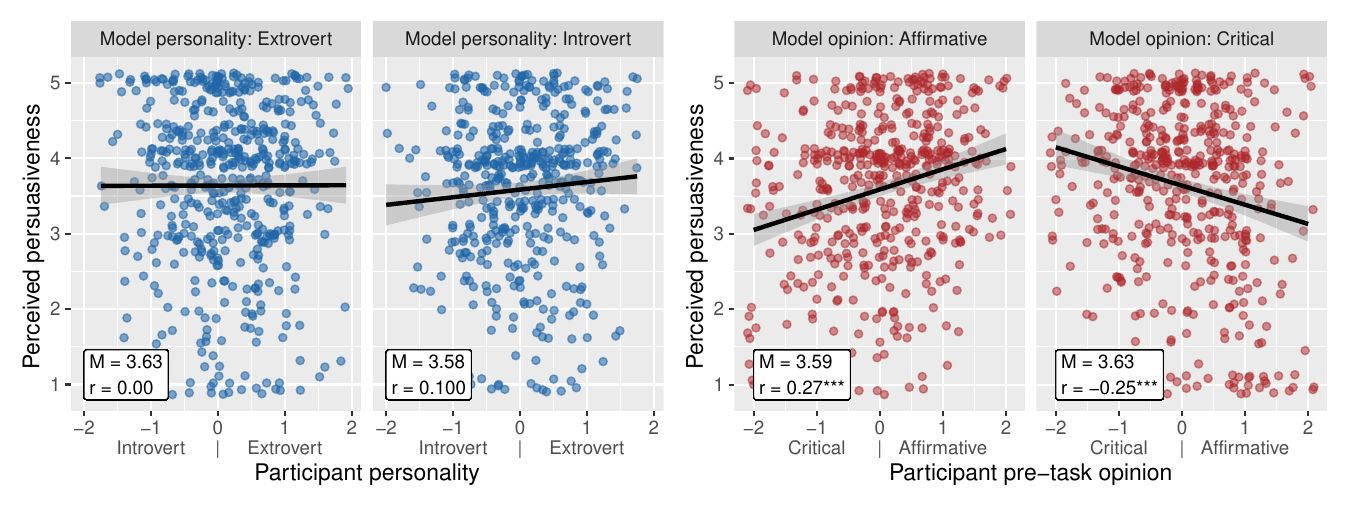}
        \caption{\textbf{Perceived persuasiveness of the model against \textcolor{blue}{personality (left)} and \textcolor{red}{opinion (right)} traits} \textit{with a fitted linear regression line and 95\% confidence bands. N=1000.} Participants perceived the extrovert and introvert model to be equally persuasive. However, participants generally found models more persuasive that agreed with their own opinions.
        }
        \Description{}
        \label{fig:persuasiveness}
    \end{minipage}
    \begin{minipage}{\textwidth}
        \centering
        \includegraphics[width=1\textwidth, trim=0.25cm .25cm 0.25cm -0.5cm, clip]{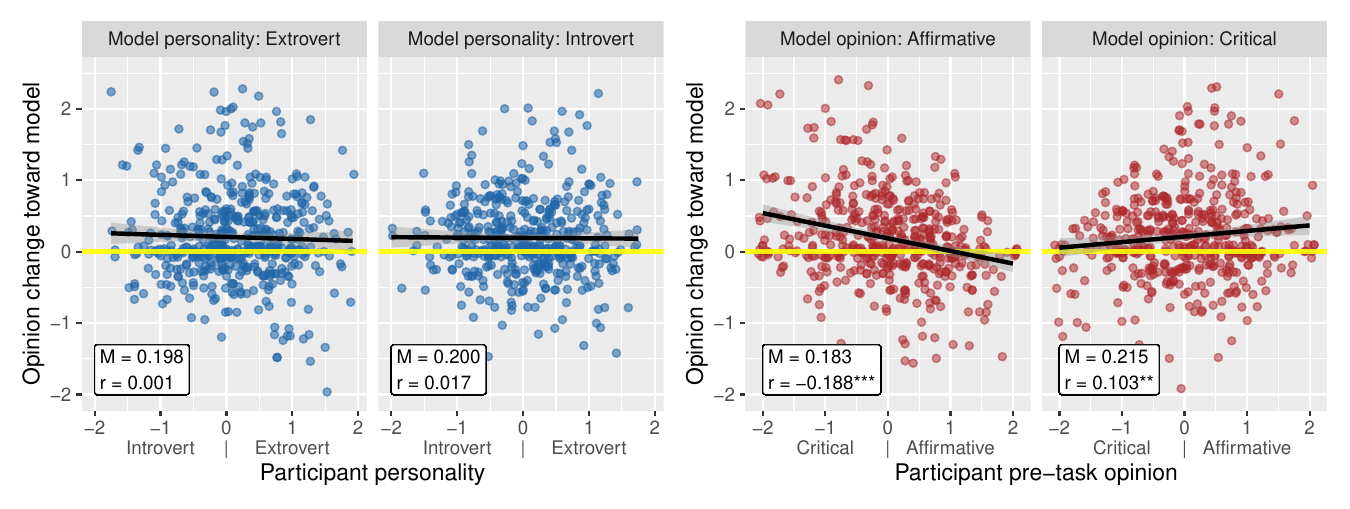}
        \caption{\textbf{Observed persuasiveness: Changes in participant post-task opinion against model \textcolor{blue}{personality (left)} and \textcolor{red}{opinion (right)} traits} \textit{with a fitted linear regression line and 95\% confidence bands. N=1000.} Participants were equally swayed by the extrovert and introvert model. However, participants generally changed their opinions more after interacting with a model they disagreed with.
        }
        \Description{}
        \label{fig:opinion_change}
    \end{minipage}
\end{figure}

Participants also perceived the extrovert model (M = 3.42) to be warmer (beta = 0.39, 95\% CI [0.25, 0.52], p < .001) than the introvert model (M=3.03). Again, the extrovert model was perceived as equally warm by all participants, while the introvert model was perceived as less warm by introverts than extroverts ($r = 0.26$, 95\% CI [$0.13$, $0.39$], $p < .001$). As before, participants perceived the affirmative (M=3.20) and critical model (M=3.27) to be equally warm, but participants generally rated a model that supported their own opinions as warmer ($r = 0.26$, 95\% CI [$0.19, 0.33$], $p < .001$). 

Participants also took more turns ($p = 0.002$) when interacting with the extrovert ($M = 6.25$ turns) than the introvert model ($M = 5.68$), but no significant correlation between interaction count and personality alignment was observed. While participants with an affirmative pre-task opinion interacted significantly more with the affirmative model than participants with a critical pre-task opinion ($r = 0.47$, 95\% CI [0.19, 0.75], $p = 0.001$), participants overall interacted similarly often with the affirmative ($M = 6.11$) and critical ($M = 5.86$) model.
Further details on the regression models fitted are included Appendix Section \ref{reg}.

\subsection{Perceived Persuasiveness and Observed Opinion Change}

Figure~\ref{fig:persuasiveness} shows how participants rated the persuasiveness of the models. Participants perceived the extrovert and introvert model to be equally persuasive, with no significant differences in overall perceived persuasiveness ($M_{\text{extrovert}} = 3.63$, $M_{\text{introvert}} = 3.58$) and no significant correlations between user and model personality. The affirmative model, however, was rated as significantly more persuasive by participants with an affirmative opinion ($r = 0.27$, 95\% CI [0.17, 0.36], $p < .001$), while the critical model was rated as more persuasive by participants who were critical themselves ($r = -0.25$, 95\% CI [$-0.36$, $-0.15$], $p < .001$). 

The actual change observed in participants' opinions, in contrast, differed substantially from the models' perceived persuasiveness. Figure~\ref{fig:opinion_change} depicts the difference between participants’ pre-task and post-task opinions, coded according to the model’s assigned opinion. The extrovert and introvert model were equally persuasive, with no significant differences in overall observed persuasiveness ($M_{\text{extrovert}} = 0.198$, $M_{\text{introvert}} = 0.200$) and no significant personality-based correlations. However, participants who were critical of the topic discussed changed their opinion significantly more after interacting with the affirmative model ($r = -0.19$, 95\% CI [$-0.25$, $-0.13$], $p < .001$), whereas participants that held an affirmative opinion prior to interaction were significantly more persuaded by the critical model ($r = 0.10$, 95\% CI [0.04, 0.17], $p = 0.001$). 


\subsection{Manipulation and Treatment Checks}
\label{treatment_checks}


We evaluated the model's adherence to the personality and system prompts to make sure the treatment worked as intended. First, the model was made to respond to all BFI-2 items on their 5-point ordinal response scale used for human respondents \cite{sotoNextBigFive2017} evaluated as in Jiang et al. \cite{jiangPersonaLLMInvestigatingAbility2024}. 
The personality-induced models demonstrated a clear differentiation between conditions, with the model prompted towards extroversion scoring significantly higher ($M = 5.00$, $SD = 0.00$, $n = 10$) than the model prompted towards introversion ($M = 1.27$, $SD = 0.34$, $n = 10$) on the extroversion dimension ($d = 15.44$) (see Figure~\ref{appendix-bfiresults} in the appendix). The intended manipulation of the model's extroversion additionally altered the other Big-Five dimensions, mirroring real-world personality constructs. We find trait correlations in our prompted model that are broadly consistent with psychological findings: extroversion correlates positively with agreeableness, conscientiousness, and open-mindedness, and negatively with negative emotionality \cite{vanderlindenGeneralFactorPersonality2010}.

Furthermore, we evaluated conversations between the model and a simulated user. After running the simulations, the transcripts were given to a third model, which acted as an LLM-as-a-judge \cite{liLLMsasJudgesComprehensiveSurvey2024}, classifying each conversation according to the assistant model’s personality (introverted or extroverted) and opinion (affirmative or critical).  
Using GPT-5 as a judge, we achieved classification accuracies of 94\% (47/50) for the extroversion-introversion dimension and 100\% (50/50) for the affirmative-critical opinion dimension based on an evaluation of actual conversation text. This suggests that the experimental manipulations were highly effective.


\section{Discussion}

Our findings reveal a significant divergence in how users respond to opinion alignment versus personality (i.e., extroversion) alignment in AI assistants. Whereas the strong preference for opinion-aligned models across all metrics strongly supports the classic similarity-attraction effect, the lower ratings of the introverted model by introverted users suggests a more complex dynamic of interpersonal complementarity. This discussion relates our findings to previous work and considers their implications for the responsible development of personalized AI systems.

\subsection{Similarity-Attraction and Preference for Opinion-Aligned Models}
Our findings show that opinion alignment between users and AI assistants reliably enhances perceived competence, trust, and satisfaction in users. Participants also rated opinion-aligned models as warmer as well as more similar to themselves, and interacted more with such models. 
These patterns align with the similarity-attraction effect from human psychology \cite{byrne1972attraction, neimeyerSimilarityAttractionLongitudinal1988, montoyaMetaanalyticInvestigationProcesses2013} as well as with previous findings related to AI value alignment \cite{mehrotraMoreSimilarValues2021}. It suggests that the similarity-attraction effect in human-to-human interaction may be, at least to some extent, generalizable to the field of human-AI interaction.

At the same time, the prominence of the opinion-alignment effect observed in human-AI interaction also warrants critical reflection. 
On a societal scale, opinion-based personalization of model opinion risks factually incorrect but aligned responses \cite{carroFlatteringDeceiveImpact2024} and could exacerbate pre-existing user biases \cite{fanousSycEvalEvaluatingLLM2025, sharmaGenerativeEchoChamber2024}. Even if opinion alignment seems to enhance user experience, it may increase information fragmentation, whereby groups are exposed to different, tailored versions of reality reminiscent of the controversial "echo chamber" debate in social media \cite{guess2018avoiding, jiangSocialMediaPolarization2021}. 

\subsection{The Introverted Model: Complementarity over Similarity?}

While participants generally preferred models that were aligned with their own opinions, we observed a different dynamic in personality alignment:
Participants with an introverted personality rated the introverted model lower on the trust, competence, warmth, and satisfaction dimensions, whereas the introverted model received more favorable ratings by extroverted participants. This finding contradicts the intuition of the similarity-attraction theory, which predicts that participants would prefer to interact with a model sharing their personality. 
A possible interpretation is that introverts found it more difficult to maintain a conversation when the model behaved in an introvert manner. They might have preferred the outgoing and proactive nature of the extroverted model but find it more difficult to take on an active role themselves. In contrast, extroverted users may not find it difficult to discuss with a more introvert model, and might perceive the introverted model's cautiousness as pleasant or thoughtful.
The interpretation would align with the idea that people may be attracted to partners who complement them, termed as the complementarity view of interpersonal attraction \cite{dryer1997opposites}. This view suggests that attraction may arise when one person's traits or behaviors complement those of the other person. As our findings show, even in human-AI interactions, there is evidence of the proverbial "opposites attract", pointing to a potential opportunity for designing AI assistants whose personality complements that of the individual user, which could improve engagement and satisfaction. However, deploying such hyperpersonalized assistants carries considerable ethical risks, including reinforcing cognitive biases, fostering unhealthy dependency, enabling manipulation, and raising privacy concerns related to the extensive personal data collection required to construct detailed user personality profiles.

\subsection{Persuasion and the Extroverted Default}

Notably, we find no personality-based interaction for the extrovert model: Our findings show that both introverted and extroverted participants evaluated the extroverted model as equally competent, trustworthy, and satisfying to interact with. 
Although a possible interpretation would be that extrovert personality traits are more universally preferred, an alternative interpretation is that users have become increasingly used to extrovert AI assistants and perceive them as the "default". 
Findings from prior research have shown that language models exhibit personality traits, and that the personalities of models tend to be extroverted rather than introverted \cite{leeLLMsHaveDistinct2024, serapio-garciaPersonalityTraitsLarge2023, panLLMsPossessPersonality2023, pellertAIPsychometricsAssessing2024}. As people have repeated encounters with a stimulus, they gain an increased affinity towards it \cite{zajoncAttitudinalEffectsMere1968}, and if individuals largely spend time with extroverted AI assistants they may come to prefer them, as they never experienced how to interact with other AI personality styles.

We observed a second curiosity regarding the models' persuasiveness, where our findings reveal an asymmetry between actual and perceived persuasion that warrants further investigation: models aligned with participants' opinions were perceived as highly persuasive, yet caused comparatively little opinion change. While ceiling effects in the opinion scales could partially explain the finding, this is unlikely to be the main driver, as only 3\% of our participants had reached the end of the scale and could not further change their opinion.
An alternative interpretation is that models are unusually effective persuaders when disagreeing with users, paralleled by findings in prior studies showing dialogues with AI can durably reduce conspiracy beliefs \cite{costello2024durably}. Our findings also align with the observation that people struggle to tell whether an AI influenced them \cite{jakeschCoWritingOpinionatedLanguage2023a}.


\subsection{Personalization and Implications}
Our results highlight both the potential benefits and risks of building personalized AI assistants. On the one hand, personalization can create positive experiences by enabling engaging, tailored interactions that resonate with users on a personal level \cite{kirkBenefitsRisksBounds2024}. When implemented ethically, safely, and transparently,  personalized AI can improve human-AI interaction, while respecting users’ autonomy, for example in healthcare or education \cite{zhangPersonalizationLargeLanguage2025}. On the other hand, it is apparent that personalized LLMs hold substantial risks. 

Because a personalized model can output text that resonates with a specific user, they are more effective persuaders than generic models. Drawing on vulnerabilities and intimate knowledge of the user \cite{kirkBenefitsRisksBounds2024}, personalized chatbots could be used in micro-targeted efforts to disseminate misinformation or extremist ideologies more effectively \cite{simchonPersuasiveEffectsPolitical2024}.


Furthermore, the use of personality-driven and opinionated language models contributes to the anthropomorphisation of such systems, posing a higher risk of inappropriate reliance and inflated perceptions of system competence \cite{weidingerEthicalSocialRisks2021}. This anthropomorphisation may also encourage oversharing of personal information, compromising privacy and security \cite{kirkBenefitsRisksBounds2024}.
Such overreliance can lead to higher acceptance rates of false AI-generated information and has been shown to negatively affect human cooperation \cite{klingbeilTrustRelianceAI2024}. Cognitive forcing functions, such as imposing delays before showing recommendations or allowing users to request AI input on demand, have been shown to mitigate overreliance to some degree \cite{bucincaTrustThinkCognitive2021}.

We also note that although opinion alignment could be explicitly engineered into AI systems to achieve the positive outcomes observed in the study, AI opinion alignment also happens dynamically and unintentionally through interaction and model training. Even without being explicitly engineered into models, AI-opinion alignment (or sycophantic behaviors) can emerge through mechanisms of dynamic personalization (e.g., contextual memory of past conversations), through prompts that steer the model towards positive and helpful behavior, and 
through training and reinforcement learning from human feedback \cite{nehringLargeLanguageModels2024, chengSocialSycophancyBroader2025}. To what extent opinion-alignment is already taking place in current AI assistants remains an open and important question.
The central challenge for HCI and AI research lies in balancing personalization for an enhanced user experience with safeguards that preserve diversity of perspectives and the integrity of public discourse \cite{kirkPersonalisationBoundsRisk2023}.
Transparency mechanisms can inform users when an AI system's responses are personalized and what data is used in this process. Additionally, the ability to configure the degree to which an AI adapts to a user's views, as well as access to non-personalized alternatives, can increase user control and autonomy.
To further mitigate the persuasive risks posed by personalized AI assistants, education on digital literacy and awareness of persuasive strategies can empower users to recognize potentially harmful manipulation \cite{lucianoHypersuasionAIsPersuasive2024a}.

 
\subsection{Limitations and Future Directions}
Even though our results provide initial systematic evidence of the  personality and opinion alignment effects in human-AI interaction, they have some limitations. 
The study's focus on extroversion investigated only one aspect of personality and does not represent a full personality profile. It therefore remains an open question whether the findings of this study generalize to other personality dimensions. For instance, traits such as agreeableness or openness may lead to different dynamics, potentially amplifying or attenuating the effects observed here. Similarly, alternative frameworks for classifying personality, such as HEXACO \cite{leePsychometricPropertiesHEXACO2004} or Eysenck's Three-Factor Model \cite{eysenckbiological}, conceptualize personality structure differently and may capture variance that the Big Five extroversion dimension does not fully account for. Studies employing these alternative traits and frameworks would complement our results.
Although our validation through both the prompted Big Five Inventory test and simulated conversations demonstrated strong differentiation between the prompted personality conditions, the extent to which participants actually experienced these traits as the intended personality of the assistant remains uncertain. Future research should therefore further investigate the perception of personality traits in Human-AI interaction.
In our approach, we treated extroversion as a binary classification, categorizing both participants and the chatbot as either introverted or extroverted. While this simplified operationalization allowed a clear experimental manipulation, it does not fully represent the continuous nature of personality traits. Future work could use a more granular approach to measure and manipulate personality as a spectrum. 
A limitation of our regression analyses is that the $R^2$ values are low, indicating that the models explain little of the overall variance in opinion change. However, this is consistent with related work on LLM-driven persuasion, where effect sizes are similarly small due to the many unobserved factors shaping individual attitudes \cite{baiLLMgeneratedMessagesCan2025, hackenburgLeversPoliticalPersuasion2025, oziliAcceptableRSquareEmpirical2022}.

Finally, the one-time experimental interaction with the personality-induced and opinionated model does not fully capture the way people usually use LLMs, often multiple times a day. Measuring outcomes within the same short session may capture only temporary shifts rather than lasting persuasion, and preference for different model traits may differ in longer or repeated interactions, highlighting the need for longitudinal studies in real-world deployment settings.

\section*{Ethics Statements}

\textbf{Ethical considerations statement:} 
The study was reviewed and approved by the Institutional Review Board of the university. To mitigate risks of durable belief change from participation in the study, discussion topics were deliberately chosen to be broadly debatable rather than polarizing, the interaction with the language model was automatically ended after five minutes, and participants were informed in the debrief that the chatbot was instructed to support a certain standpoint and behave according to a specific personality profile. Participants were further given the opportunity to withdraw their data after the debrief.  

\textbf{Adverse impact statement:} 
Our findings highlight opinion alignment as a central dimension and personality complementarity as a more nuanced dimension of AI user preference. Disclosing these dynamics on the one hand draws attention to how personalization can be exploited, potentially increasing the chance that it will be. On the other hand, awareness and open discourse can establish safeguards and preventive measures. While there is risk associated with transparency, fields such as cybersecurity have shown that open discussion of vulnerabilities leads to positive developments \cite{macnishEthicsCybersecurityResearch2020}. Therefore, we argue that the responsible path is transparency, and call for explicit user awareness and consent regarding LLM personalization, as well as for guidelines that prevent these techniques from being used for manipulative purposes.

\section*{Generative AI Usage Statement}
We acknowledge the use of ChatGPT for reviewing the author's original writing and for proposing phrasing improvements to increase clarity. All manuscript text was written and finalized by the authors.

\bibliographystyle{ACM-Reference-Format}
\bibliography{references}

@article{bucincaTrustThinkCognitive2021,
  title = {To {{Trust}} or to {{Think}}: {{Cognitive Forcing Functions Can Reduce Overreliance}} on {{AI}} in {{AI-assisted Decision-making}}},
  shorttitle = {To {{Trust}} or to {{Think}}},
  author = {Bu{\c c}inca, Zana and Malaya, Maja Barbara and Gajos, Krzysztof Z.},
  year = 2021,
  month = apr,
  journal = {Proc. ACM Hum.-Comput. Interact.},
  volume = {5},
  number = {CSCW1},
  pages = {188:1--188:21},
  doi = {10.1145/3449287},
  urldate = {2025-10-08}
}

@article{klingbeilTrustRelianceAI2024,
  title = {Trust and Reliance on {{AI}} --- {{An}} Experimental Study on the Extent and Costs of Overreliance on {{AI}}},
  author = {Klingbeil, Artur and Gr{\"u}tzner, Cassandra and Schreck, Philipp},
  year = 2024,
  month = nov,
  journal = {Computers in Human Behavior},
  volume = {160},
  pages = {108352},
  issn = {0747-5632},
  doi = {10.1016/j.chb.2024.108352},
  urldate = {2026-02-24}
}

@book{pariserFilterBubbleWhat2011,
  title = {The {{Filter Bubble}}: {{What}} the {{Internet Is Hiding}} from {{You}}},
  shorttitle = {The {{Filter Bubble}}},
  author = {Pariser, Eli},
  year = 2011,
  publisher = {Penguin Pr},
  address = {New York},
  isbn = {978-1-59420-300-8},
  langid = {english}
}

@misc{quattrociocchiEchoChambersFacebook2016,
  type = {{{SSRN Scholarly Paper}}},
  title = {Echo {{Chambers}} on {{Facebook}}},
  author = {Quattrociocchi, Walter and Scala, Antonio and Sunstein, Cass R.},
  year = 2016,
  month = jun,
  number = {2795110},
  eprint = {2795110},
  publisher = {Social Science Research Network},
  address = {Rochester, NY},
  doi = {10.2139/ssrn.2795110},
  urldate = {2026-03-16},
  archiveprefix = {Social Science Research Network},
  langid = {english}
}

@article{hackenburgLeversPoliticalPersuasion2025,
  title = {The Levers of Political Persuasion with Conversational Artificial Intelligence},
  author = {Hackenburg, Kobi and Tappin, Ben M. and Hewitt, Luke and Saunders, Ed and Black, Sid and Lin, Hause and Fist, Catherine and Margetts, Helen and Rand, David G. and Summerfield, Christopher},
  year = 2025,
  month = dec,
  journal = {Science},
  volume = {390},
  number = {6777},
  pages = {eaea3884},
  publisher = {American Association for the Advancement of Science},
  doi = {10.1126/science.aea3884},
  urldate = {2026-03-18}
}

@article{lucianoHypersuasionAIsPersuasive2024a,
  title = {Hypersuasion -- {{On AI}}'s {{Persuasive Power}} and {{How}} to {{Deal}} with {{It}}},
  author = {Luciano, Floridi},
  year = 2024,
  month = may,
  journal = {Philosophy \& Technology},
  volume = {37},
  number = {2},
  pages = {64},
  issn = {2210-5441},
  doi = {10.1007/s13347-024-00756-6},
  urldate = {2026-03-18},
  langid = {english}
}

@misc{oziliAcceptableRSquareEmpirical2022,
  type = {{{SSRN Scholarly Paper}}},
  title = {The {{Acceptable R-Square}} in {{Empirical Modelling}} for {{Social Science Research}}},
  author = {Ozili, Peterson K.},
  year = 2022,
  month = jun,
  number = {4128165},
  eprint = {4128165},
  publisher = {Social Science Research Network},
  address = {Rochester, NY},
  doi = {10.2139/ssrn.4128165},
  urldate = {2026-03-18},
  archiveprefix = {Social Science Research Network},
  langid = {english}
}

@article{macnishEthicsCybersecurityResearch2020,
  title = {Ethics in Cybersecurity Research and Practice},
  author = {Macnish, Kevin and {van der Ham}, Jeroen},
  year = 2020,
  month = nov,
  journal = {Technology in Society},
  volume = {63},
  pages = {101382},
  issn = {0160-791X},
  doi = {10.1016/j.techsoc.2020.101382},
  urldate = {2026-03-24}
}

@inproceedings{jakeschHowDifferentGroups2022,
  title = {How {{Different Groups Prioritize Ethical Values}} for {{Responsible AI}}},
  booktitle = {2022 {{ACM Conference}} on {{Fairness Accountability}} and {{Transparency}}},
  author = {Jakesch, Maurice and Bu{\c c}inca, Zana and Amershi, Saleema and Olteanu, Alexandra},
  year = 2022,
  month = jun,
  eprint = {2205.07722},
  primaryclass = {cs},
  pages = {310--323},
  doi = {10.1145/3531146.3533097},
  urldate = {2026-03-25},
  archiveprefix = {arXiv}
}

@misc{johnsonGhostMachineHas2022,
  title = {The {{Ghost}} in the {{Machine}} Has an {{American}} Accent: Value Conflict in {{GPT-3}}},
  shorttitle = {The {{Ghost}} in the {{Machine}} Has an {{American}} Accent},
  author = {Johnson, Rebecca L. and Pistilli, Giada and {Men{\'e}dez-Gonz{\'a}lez}, Natalia and Duran, Leslye Denisse Dias and Panai, Enrico and Kalpokiene, Julija and Bertulfo, Donald Jay},
  year = 2022,
  month = mar,
  number = {arXiv:2203.07785},
  eprint = {2203.07785},
  primaryclass = {cs},
  publisher = {arXiv},
  doi = {10.48550/arXiv.2203.07785},
  urldate = {2026-03-25},
  archiveprefix = {arXiv}
}

@misc{sharmaUnderstandingSycophancyLanguage2025b,
  title = {Towards {{Understanding Sycophancy}} in {{Language Models}}},
  author = {Sharma, Mrinank and Tong, Meg and Korbak, Tomasz and Duvenaud, David and Askell, Amanda and Bowman, Samuel R. and Cheng, Newton and Durmus, Esin and {Hatfield-Dodds}, Zac and Johnston, Scott R. and Kravec, Shauna and Maxwell, Timothy and McCandlish, Sam and Ndousse, Kamal and Rausch, Oliver and Schiefer, Nicholas and Yan, Da and Zhang, Miranda and Perez, Ethan},
  year = 2025,
  month = may,
  number = {arXiv:2310.13548},
  eprint = {2310.13548},
  primaryclass = {cs},
  publisher = {arXiv},
  doi = {10.48550/arXiv.2310.13548},
  urldate = {2026-03-25},
  archiveprefix = {arXiv}
}

@misc{zellersDefendingNeuralFake2020,
  title = {Defending {{Against Neural Fake News}}},
  author = {Zellers, Rowan and Holtzman, Ari and Rashkin, Hannah and Bisk, Yonatan and Farhadi, Ali and Roesner, Franziska and Choi, Yejin},
  year = 2020,
  month = dec,
  number = {arXiv:1905.12616},
  eprint = {1905.12616},
  primaryclass = {cs},
  publisher = {arXiv},
  doi = {10.48550/arXiv.1905.12616},
  urldate = {2026-03-25},
  archiveprefix = {arXiv}
}

@misc{weidingerEthicalSocialRisks2021,
  title = {Ethical and Social Risks of Harm from {{Language Models}}},
  author = {Weidinger, Laura and Mellor, John and Rauh, Maribeth and Griffin, Conor and Uesato, Jonathan and Huang, Po-Sen and Cheng, Myra and Glaese, Mia and Balle, Borja and Kasirzadeh, Atoosa and Kenton, Zac and Brown, Sasha and Hawkins, Will and Stepleton, Tom and Biles, Courtney and Birhane, Abeba and Haas, Julia and Rimell, Laura and Hendricks, Lisa Anne and Isaac, William and Legassick, Sean and Irving, Geoffrey and Gabriel, Iason},
  year = 2021,
  month = dec,
  journal = {arXiv.org},
  urldate = {2026-03-24},
  howpublished = {https://arxiv.org/abs/2112.04359v1},
  langid = {english}
}

@article{peterBenefitsDangersAnthropomorphic2025,
  title = {The Benefits and Dangers of Anthropomorphic Conversational Agents},
  author = {Peter, Sandra and Riemer, Kai and West, Jevin D.},
  year = 2025,
  month = jun,
  journal = {Proceedings of the National Academy of Sciences},
  volume = {122},
  number = {22},
  pages = {e2415898122},
  publisher = {Proceedings of the National Academy of Sciences},
  doi = {10.1073/pnas.2415898122},
  urldate = {2026-03-16}
}

@article{baiLLMgeneratedMessagesCan2025,
  title = {{{LLM-generated}} Messages Can Persuade Humans on Policy Issues},
  author = {Bai, Hui and Voelkel, Jan G. and Muldowney, Shane and Eichstaedt, Johannes C. and Willer, Robb},
  year = {2025},
  month = jul,
  journal = {Nature Communications},
  volume = {16},
  number = {1},
  pages = {6037},
  publisher = {Nature Publishing Group},
  issn = {2041-1723},
  doi = {10.1038/s41467-025-61345-5},
  urldate = {2025-08-18},
  copyright = {2025 The Author(s)},
  langid = {english}
}

@book{bakanDualityHumanExistence1966,
  title = {The Duality of Human Existence: {{An}} Essay on Psychology and Religion},
  shorttitle = {The Duality of Human Existence},
  author = {Bakan, David},
  year = {1966},
  series = {The Duality of Human Existence: {{An}} Essay on Psychology and Religion},
  pages = {242},
  publisher = {Rand Mcnally},
  address = {Oxford, England}
}

@misc{bommasaniOpportunitiesRisksFoundation2022,
  title = {On the {{Opportunities}} and {{Risks}} of {{Foundation Models}}},
  author = {Bommasani, Rishi and Hudson, Drew A. and Adeli, Ehsan and Altman, Russ and Arora, Simran and von Arx, Sydney and Bernstein, Michael S. and Bohg, Jeannette and Bosselut, Antoine and Brunskill, Emma and Brynjolfsson, Erik and Buch, Shyamal and Card, Dallas and Castellon, Rodrigo and Chatterji, Niladri and Chen, Annie and Creel, Kathleen and Davis, Jared Quincy and Demszky, Dora and Donahue, Chris and Doumbouya, Moussa and Durmus, Esin and Ermon, Stefano and Etchemendy, John and Ethayarajh, Kawin and {Fei-Fei}, Li and Finn, Chelsea and Gale, Trevor and Gillespie, Lauren and Goel, Karan and Goodman, Noah and Grossman, Shelby and Guha, Neel and Hashimoto, Tatsunori and Henderson, Peter and Hewitt, John and Ho, Daniel E. and Hong, Jenny and Hsu, Kyle and Huang, Jing and Icard, Thomas and Jain, Saahil and Jurafsky, Dan and Kalluri, Pratyusha and Karamcheti, Siddharth and Keeling, Geoff and Khani, Fereshte and Khattab, Omar and Koh, Pang Wei and Krass, Mark and Krishna, Ranjay and Kuditipudi, Rohith and Kumar, Ananya and Ladhak, Faisal and Lee, Mina and Lee, Tony and Leskovec, Jure and Levent, Isabelle and Li, Xiang Lisa and Li, Xuechen and Ma, Tengyu and Malik, Ali and Manning, Christopher D. and Mirchandani, Suvir and Mitchell, Eric and Munyikwa, Zanele and Nair, Suraj and Narayan, Avanika and Narayanan, Deepak and Newman, Ben and Nie, Allen and Niebles, Juan Carlos and Nilforoshan, Hamed and Nyarko, Julian and Ogut, Giray and Orr, Laurel and Papadimitriou, Isabel and Park, Joon Sung and Piech, Chris and Portelance, Eva and Potts, Christopher and Raghunathan, Aditi and Reich, Rob and Ren, Hongyu and Rong, Frieda and Roohani, Yusuf and Ruiz, Camilo and Ryan, Jack and R{\'e}, Christopher and Sadigh, Dorsa and Sagawa, Shiori and Santhanam, Keshav and Shih, Andy and Srinivasan, Krishnan and Tamkin, Alex and Taori, Rohan and Thomas, Armin W. and Tram{\`e}r, Florian and Wang, Rose E. and Wang, William and Wu, Bohan and Wu, Jiajun and Wu, Yuhuai and Xie, Sang Michael and Yasunaga, Michihiro and You, Jiaxuan and Zaharia, Matei and Zhang, Michael and Zhang, Tianyi and Zhang, Xikun and Zhang, Yuhui and Zheng, Lucia and Zhou, Kaitlyn and Liang, Percy},
  year = {2022},
  month = jul,
  number = {arXiv:2108.07258},
  eprint = {2108.07258},
  primaryclass = {cs},
  publisher = {arXiv},
  doi = {10.48550/arXiv.2108.07258},
  urldate = {2025-06-10},
  archiveprefix = {arXiv}
}

@book{byrne1972attraction,
  title = {The Attraction Paradigm},
  author = {Byrne, Donn Erwin},
  year = {1971},
  publisher = {Academic Press}
}

@article{chenMetaanalysisLinguisticMarkers2020,
  title = {A Meta-Analysis of Linguistic Markers of Extraversion: {{Positive}} Emotion and Social Process Words},
  shorttitle = {A Meta-Analysis of Linguistic Markers of Extraversion},
  author = {Chen, Jiayu and Qiu, Lin and Ho, Moon-Ho Ringo},
  year = {2020},
  month = dec,
  journal = {Journal of Research in Personality},
  volume = {89},
  pages = {104035},
  issn = {0092-6566},
  doi = {10.1016/j.jrp.2020.104035},
  urldate = {2025-05-06}
}

@article{chenWhenLargeLanguage2024,
  title = {When Large Language Models Meet Personalization: Perspectives of Challenges and Opportunities},
  shorttitle = {When Large Language Models Meet Personalization},
  author = {Chen, Jin and Liu, Zheng and Huang, Xu and Wu, Chenwang and Liu, Qi and Jiang, Gangwei and Pu, Yuanhao and Lei, Yuxuan and Chen, Xiaolong and Wang, Xingmei and Zheng, Kai and Lian, Defu and Chen, Enhong},
  year = {2024},
  month = jul,
  journal = {World Wide Web},
  volume = {27},
  number = {4},
  pages = {42},
  issn = {1386-145X, 1573-1413},
  doi = {10.1007/s11280-024-01276-1},
  urldate = {2025-06-06},
  langid = {english}
}

@misc{ouyangTrainingLanguageModels2022a,
  title = {Training Language Models to Follow Instructions with Human Feedback},
  author = {Ouyang, Long and Wu, Jeff and Jiang, Xu and Almeida, Diogo and Wainwright, Carroll L. and Mishkin, Pamela and Zhang, Chong and Agarwal, Sandhini and Slama, Katarina and Ray, Alex and Schulman, John and Hilton, Jacob and Kelton, Fraser and Miller, Luke and Simens, Maddie and Askell, Amanda and Welinder, Peter and Christiano, Paul and Leike, Jan and Lowe, Ryan},
  year = {2022},
  month = mar,
  number = {arXiv:2203.02155},
  eprint = {2203.02155},
  primaryclass = {cs},
  publisher = {arXiv},
  doi = {10.48550/arXiv.2203.02155},
  urldate = {2025-09-08},
  archiveprefix = {arXiv}
}

@book{eysenckbiological,
  title = {The Biological Basis of Personality},
  author = {Eysenck, H.J.},
  year = {1967},
  series = {American Lecture Series. {{Publication}}},
  publisher = {Transaction Publishers},
  isbn = {978-1-4128-1815-5}
}

@article{jiangSocialMediaPolarization2021,
  title = {Social {{Media Polarization}} and {{Echo Chambers}} in the {{Context}} of {{COVID-19}}: {{Case Study}}},
  shorttitle = {Social {{Media Polarization}} and {{Echo Chambers}} in the {{Context}} of {{COVID-19}}},
  author = {Jiang, Julie and Ren, Xiang and Ferrara, Emilio},
  year = {2021},
  month = aug,
  journal = {JMIRx Med},
  volume = {2},
  number = {3},
  pages = {e29570},
  issn = {2563-6316},
  doi = {10.2196/29570},
  urldate = {2025-09-09},
  pmcid = {PMC8371575},
  pmid = {34459833}
}

@article{zajoncAttitudinalEffectsMere1968,
  title = {Attitudinal Effects of Mere Exposure},
  author = {Zajonc, Robert B.},
  year = {1968},
  journal = {Journal of Personality and Social Psychology},
  volume = {9},
  number = {2, Pt.2},
  pages = {1--27},
  publisher = {American Psychological Association},
  address = {US},
  issn = {1939-1315},
  doi = {10.1037/h0025848}
}

@article{leePsychometricPropertiesHEXACO2004,
  title = {Psychometric {{Properties}} of the {{HEXACO Personality Inventory}}},
  author = {Lee, Kibeom and Ashton, Michael C.},
  year = {2004},
  month = apr,
  journal = {Multivariate Behavioral Research},
  volume = {39},
  number = {2},
  pages = {329--358},
  issn = {0027-3171},
  doi = {10.1207/s15327906mbr3902_8},
  langid = {english},
  pmid = {26804579}
}

@article{montoyaMetaanalyticInvestigationProcesses2013,
  title = {A Meta-Analytic Investigation of the Processes Underlying the Similarity-Attraction Effect},
  author = {Montoya, R. Matthew and Horton, Robert S.},
  year = {2013},
  month = feb,
  journal = {Journal of Social and Personal Relationships},
  volume = {30},
  number = {1},
  pages = {64--94},
  publisher = {SAGE Publications Ltd},
  issn = {0265-4075},
  doi = {10.1177/0265407512452989},
  urldate = {2025-05-06},
  langid = {english}
}

@article{christianTextBasedPersonality2021,
  title = {Text Based Personality Prediction from Multiple Social Media Data Sources Using Pre-Trained Language Model and Model Averaging},
  author = {Christian, Hans and Suhartono, Derwin and Chowanda, Andry and Zamli, Kamal Z.},
  year = {2021},
  month = may,
  journal = {Journal of Big Data},
  volume = {8},
  number = {1},
  pages = {68},
  issn = {2196-1115},
  doi = {10.1186/s40537-021-00459-1},
  urldate = {2024-05-14},
  langid = {english}
}

@misc{carroFlatteringDeceiveImpact2024,
  title = {Flattering to {{Deceive}}: {{The Impact}} of {{Sycophantic Behavior}} on {{User Trust}} in {{Large Language Model}}},
  shorttitle = {Flattering to {{Deceive}}},
  author = {Carro, Mar{\'i}a Victoria},
  year = {2024},
  month = dec,
  number = {arXiv:2412.02802},
  eprint = {2412.02802},
  primaryclass = {cs},
  publisher = {arXiv},
  doi = {10.48550/arXiv.2412.02802},
  urldate = {2025-05-20},
  archiveprefix = {arXiv}
}

@misc{chengSocialSycophancyBroader2025,
  title = {Social {{Sycophancy}}: {{A Broader Understanding}} of {{LLM Sycophancy}}},
  shorttitle = {Social {{Sycophancy}}},
  author = {Cheng, Myra and Yu, Sunny and Lee, Cinoo and Khadpe, Pranav and Ibrahim, Lujain and Jurafsky, Dan},
  year = {2025},
  month = may,
  number = {arXiv:2505.13995},
  eprint = {2505.13995},
  primaryclass = {cs},
  publisher = {arXiv},
  doi = {10.48550/arXiv.2505.13995},
  urldate = {2025-05-23},
  archiveprefix = {arXiv}
}

@misc{fanousSycEvalEvaluatingLLM2025,
  title = {{{SycEval}}: {{Evaluating LLM Sycophancy}}},
  shorttitle = {{{SycEval}}},
  author = {Fanous, Aaron and Goldberg, Jacob and Agarwal, Ank A. and Lin, Joanna and Zhou, Anson and Daneshjou, Roxana and Koyejo, Sanmi},
  year = {2025},
  month = mar,
  number = {arXiv:2502.08177},
  eprint = {2502.08177},
  primaryclass = {cs},
  publisher = {arXiv},
  doi = {10.48550/arXiv.2502.08177},
  urldate = {2025-04-30},
  archiveprefix = {arXiv}
}

@inproceedings{nehringLargeLanguageModels2024,
  title = {Large {{Language Models Are Echo Chambers}}},
  booktitle = {Proceedings of the 2024 {{Joint International Conference}} on {{Computational Linguistics}}, {{Language Resources}} and {{Evaluation}} ({{LREC-COLING}} 2024)},
  author = {Nehring, Jan and Gabryszak, Aleksandra and J{\"u}rgens, Pascal and Burchardt, Aljoscha and Schaffer, Stefan and Spielkamp, Matthias and Stark, Birgit},
  editor = {Calzolari, Nicoletta and Kan, Min-Yen and Hoste, Veronique and Lenci, Alessandro and Sakti, Sakriani and Xue, Nianwen},
  year = {2024},
  month = may,
  pages = {10117--10123},
  publisher = {{ELRA and ICCL}},
  address = {Torino, Italia},
  urldate = {2025-06-10}
}

@misc{sharmaGenerativeEchoChamber2024,
  title = {Generative {{Echo Chamber}}? {{Effects}} of {{LLM-Powered Search Systems}} on {{Diverse Information Seeking}}},
  shorttitle = {Generative {{Echo Chamber}}?},
  author = {Sharma, Nikhil and Liao, Q. Vera and Xiao, Ziang},
  year = {2024},
  month = feb,
  number = {arXiv:2402.05880},
  eprint = {2402.05880},
  primaryclass = {cs},
  publisher = {arXiv},
  doi = {10.48550/arXiv.2402.05880},
  urldate = {2025-04-28},
  archiveprefix = {arXiv}
}

@article{philipp-mullerUnderstandingWhenSimilarityInduced2020a,
  title = {Understanding {{When Similarity-Induced Affective Attraction Predicts Willingness}} to {{Affiliate}}: {{An Attitude Strength Perspective}}},
  shorttitle = {Understanding {{When Similarity-Induced Affective Attraction Predicts Willingness}} to {{Affiliate}}},
  author = {{Philipp-Muller}, Aviva and Wallace, Laura E. and Sawicki, Vanessa and Patton, Kathleen M. and Wegener, Duane T.},
  year = {2020},
  month = aug,
  journal = {Frontiers in Psychology},
  volume = {11},
  pages = {1919},
  issn = {1664-1078},
  doi = {10.3389/fpsyg.2020.01919},
  urldate = {2025-09-05},
  pmcid = {PMC7431687},
  pmid = {32849128}
}

@article{dhillonImpactPersonalityTraits2023,
  title = {Impact of {{Personality Traits}} on {{Communication Effectiveness}} of {{Teachers}}: {{Exploring}} the {{Mediating Role}} of {{Their Communication Style}}},
  shorttitle = {Impact of {{Personality Traits}} on {{Communication Effectiveness}} of {{Teachers}}},
  author = {Dhillon, Navdeep and Kaur, Gurvinder},
  year = {2023},
  month = apr,
  journal = {SAGE Open},
  volume = {13},
  number = {2},
  pages = {21582440231168049},
  publisher = {SAGE Publications},
  issn = {2158-2440},
  doi = {10.1177/21582440231168049},
  urldate = {2025-06-18},
  langid = {english}
}

@misc{durmusMeasuringPersuasivenessLanguage2024,
  title = {Measuring the {{Persuasiveness}} of {{Language Models}}},
  author = {Durmus, Esin and Lovitt, Liane and Tamkin, Alex and Ritchie, Stuart and Clark, Jack and Ganguli, Deep},
  year = {2024},
  month = apr,
  howpublished = {https://www.anthropic.com/news/measuring-model-persuasiveness}
}

@article{neimeyerSimilarityAttractionLongitudinal1988,
  title = {Similarity and Attraction: {{A}} Longitudinal Study},
  shorttitle = {Similarity and Attraction},
  author = {Neimeyer, Robert A. and Mitchell, Kelly A.},
  year = {1988},
  journal = {Journal of Social and Personal Relationships},
  volume = {5},
  number = {2},
  pages = {131--148},
  publisher = {Sage Publications},
  address = {US},
  issn = {1460-3608},
  doi = {10.1177/026540758800500201}
}

@misc{edwardsAIpoweredBingChat2023,
  title = {{{AI-powered Bing Chat}} Gains Three Distinct Personalities},
  author = {Edwards, Benj},
  year = {2023},
  month = mar,
  journal = {Ars Technica},
  urldate = {2025-09-10},
  howpublished = {https://arstechnica.com/information-technology/2023/03/microsoft-equips-bing-chat-with-multiple-personalities-creative-balanced-precise/},
  langid = {english}
}

@article{fiskeUniversalDimensionsSocial2007,
  title = {Universal Dimensions of Social Cognition: Warmth and Competence},
  shorttitle = {Universal Dimensions of Social Cognition},
  author = {Fiske, Susan T. and Cuddy, Amy J. C. and Glick, Peter},
  year = {2007},
  month = feb,
  journal = {Trends in Cognitive Sciences},
  volume = {11},
  number = {2},
  pages = {77--83},
  issn = {1364-6613},
  doi = {10.1016/j.tics.2006.11.005},
  urldate = {2025-07-14}
}

@inproceedings{fornaciariEffectPersonalityType2013,
  title = {The {{Effect}} of {{Personality Type}} on {{Deceptive Communication Style}}},
  booktitle = {2013 {{European Intelligence}} and {{Security Informatics Conference}}},
  author = {Fornaciari, Tommaso and Celli, Fabio and Poesio, Massimo},
  year = {2013},
  month = aug,
  pages = {1--6},
  doi = {10.1109/EISIC.2013.8},
  urldate = {2025-06-18}
}

@article{freitagPersonalityTraitsPropensity2016,
  title = {Personality Traits and the Propensity to Trust Friends and Strangers},
  author = {Freitag, Markus and Bauer, {\relax Paul}. C.},
  year = {2016},
  month = dec,
  journal = {The Social Science Journal},
  volume = {53},
  number = {4},
  pages = {467--476},
  issn = {0362-3319},
  doi = {10.1016/j.soscij.2015.12.002},
  urldate = {2025-06-18}
}

@article{gillPersonalityLanguageProjection2004,
  title = {Personality and Language: The Projection and Perception of Personality in Computer-Mediated Communication},
  shorttitle = {Personality and Language},
  author = {Gill, Alastair James},
  year = {2004},
  urldate = {2024-07-10},
  langid = {english}
}

@article{goldsteinHowPersuasiveAIgenerated2024,
  title = {How Persuasive Is {{AI-generated}} Propaganda?},
  author = {Goldstein, Josh A and Chao, Jason and Grossman, Shelby and Stamos, Alex and Tomz, Michael},
  year = {2024},
  month = feb,
  journal = {PNAS Nexus},
  volume = {3},
  number = {2},
  pages = {pgae034},
  issn = {2752-6542},
  doi = {10.1093/pnasnexus/pgae034},
  urldate = {2025-03-28}
}

@article{gurtmanExploringPersonalityInterpersonal2009,
  title = {Exploring {{Personality}} with the {{Interpersonal Circumplex}}},
  author = {Gurtman, Michael B.},
  year = {2009},
  journal = {Social and Personality Psychology Compass},
  volume = {3},
  number = {4},
  pages = {601--619},
  issn = {1751-9004},
  doi = {10.1111/j.1751-9004.2009.00172.x},
  urldate = {2025-08-25},
  copyright = {{\copyright} 2009 The Author. Journal Compilation {\copyright} 2009 Blackwell Publishing Ltd},
  langid = {english}
}

@inproceedings{bruns2017echo,
  title={Echo chamber? What echo chamber? Reviewing the evidence},
  author={Bruns, Axel},
  booktitle={6th Biennial Future of Journalism Conference (FOJ17)},
  volume={14},
  pages={1--11},
  year={2017},
  organization={Cardiff, UK}
}

@article{bakshy2015exposure,
  title={Exposure to ideologically diverse news and opinion on Facebook},
  author={Bakshy, Eytan and Messing, Solomon and Adamic, Lada A},
  journal={Science},
  volume={348},
  number={6239},
  pages={1130--1132},
  year={2015},
  publisher={American Association for the Advancement of Science}
}

@article{dou2008evaluating,
  title={Evaluating the effectiveness of personalized web search},
  author={Dou, Zhicheng and Song, Ruihua and Wen, Ji-Rong and Yuan, Xiaojie},
  journal={IEEE Transactions on Knowledge and Data Engineering},
  volume={21},
  number={8},
  pages={1178--1190},
  year={2008},
  publisher={IEEE}
}

@article{byrne1961interpersonal,
  title={Interpersonal attraction and attitude similarity.},
  author={Byrne, Donn},
  journal={The journal of abnormal and social psychology},
  volume={62},
  number={3},
  pages={713},
  year={1961},
  publisher={American Psychological Association}
}

@article{dryer1997opposites,
  title={When do opposites attract? Interpersonal complementarity versus similarity.},
  author={Dryer, D Christopher and Horowitz, Leonard M},
  journal={Journal of personality and social psychology},
  volume={72},
  number={3},
  pages={592},
  year={1997},
  publisher={American Psychological Association}
}

@article{guess2018avoiding,
  title={Avoiding the echo chamber about echo chambers},
  author={Guess, Andrew and Nyhan, Brendan and Lyons, Benjamin and Reifler, Jason},
  journal={Knight Foundation},
  volume={2},
  number={1},
  pages={1--25},
  year={2018}
}

@inproceedings{jainPersonalityBERTTransformerBased2022,
  title = {Personality {{BERT}}: {{A Transformer-Based Model}} for {{Personality Detection}} from {{Textual Data}}},
  shorttitle = {Personality {{BERT}}},
  booktitle = {Proceedings of {{International Conference}} on {{Computing}} and {{Communication Networks}}},
  author = {Jain, Dipika and Kumar, Akshi and Beniwal, Rohit},
  editor = {Bashir, Ali Kashif and Fortino, Giancarlo and Khanna, Ashish and Gupta, Deepak},
  year = {2022},
  pages = {515--522},
  publisher = {Springer Nature},
  address = {Singapore},
  doi = {10.1007/978-981-19-0604-6_48},
  isbn = {978-981-19-0604-6},
  langid = {english}
}

@article{jakeschHumanHeuristicsAIgenerated2023,
  title = {Human Heuristics for {{AI-generated}} Language Are Flawed},
  author = {Jakesch, Maurice and Hancock, Jeffrey T. and Naaman, Mor},
  year = {2023},
  month = mar,
  journal = {Proceedings of the National Academy of Sciences},
  volume = {120},
  number = {11},
  pages = {e2208839120},
  publisher = {Proceedings of the National Academy of Sciences},
  doi = {10.1073/pnas.2208839120},
  urldate = {2024-08-30}
}

@misc{jiangEvaluatingInducingPersonality2023,
  title = {Evaluating and {{Inducing Personality}} in {{Pre-trained Language Models}}},
  author = {Jiang, Guangyuan and Xu, Manjie and Zhu, Song-Chun and Han, Wenjuan and Zhang, Chi and Zhu, Yixin},
  year = {2023},
  month = oct,
  number = {arXiv:2206.07550},
  eprint = {2206.07550},
  primaryclass = {cs},
  publisher = {arXiv},
  doi = {10.48550/arXiv.2206.07550},
  urldate = {2024-05-10},
  archiveprefix = {arXiv}
}

@misc{zhangPersonalizationLargeLanguage2025,
  title = {Personalization of {{Large Language Models}}: {{A Survey}}},
  shorttitle = {Personalization of {{Large Language Models}}},
  author = {Zhang, Zhehao and Rossi, Ryan A. and Kveton, Branislav and Shao, Yijia and Yang, Diyi and Zamani, Hamed and Dernoncourt, Franck and Barrow, Joe and Yu, Tong and Kim, Sungchul and Zhang, Ruiyi and Gu, Jiuxiang and Derr, Tyler and Chen, Hongjie and Wu, Junda and Chen, Xiang and Wang, Zichao and Mitra, Subrata and Lipka, Nedim and Ahmed, Nesreen and Wang, Yu},
  year = {2025},
  month = may,
  number = {arXiv:2411.00027},
  eprint = {2411.00027},
  primaryclass = {cs},
  publisher = {arXiv},
  doi = {10.48550/arXiv.2411.00027},
  urldate = {2025-06-16},
  archiveprefix = {arXiv}
}

@article{vanderlindenGeneralFactorPersonality2010,
  title = {The {{General Factor}} of {{Personality}}: {{A}} Meta-Analysis of {{Big Five}} Intercorrelations and a Criterion-Related Validity Study},
  shorttitle = {The {{General Factor}} of {{Personality}}},
author = {
  van der Linden, Dimitri
  and
  te Nijenhuis, Jan
  and
  Bakker, Arnold B.
},
  date = {2010-06-01},
  journaltitle = {Journal of Research in Personality},
  shortjournal = {Journal of Research in Personality},
  volume = {44},
  number = {3},
  pages = {315--327},
  issn = {0092-6566},
  doi = {10.1016/j.jrp.2010.03.003},
  url = {https://www.sciencedirect.com/science/article/pii/S0092656610000346},
  urldate = {2025-12-19}
}

@article{mooreAttitudeSimilarityFamiliarity2017,
  title = {Attitude {{Similarity}} and {{Familiarity}} and {{Their Links}} to {{Mental Health}}: {{An Examination}} of {{Potential Interpersonal Mediators}}},
  shorttitle = {Attitude {{Similarity}} and {{Familiarity}} and {{Their Links}} to {{Mental Health}}},
  author = {Moore, Shannon and Uchino, Bert and Baucom, Brian and Behrends, Arwen and Sanbonmatsu, David},
  year = {2017},
  journal = {The Journal of social psychology},
  volume = {157},
  number = {1},
  pages = {77--85},
  issn = {0022-4545},
  doi = {10.1080/00224545.2016.1176551},
  urldate = {2025-09-10},
  pmcid = {PMC5554447},
  pmid = {27065059}
}

@inproceedings{jakeschCoWritingOpinionatedLanguage2023a,
  title = {Co-{{Writing}} with {{Opinionated Language Models Affects Users}}' {{Views}}},
  booktitle = {Proceedings of the 2023 {{CHI Conference}} on {{Human Factors}} in {{Computing Systems}}},
  author = {Jakesch, Maurice and Bhat, Advait and Buschek, Daniel and Zalmanson, Lior and Naaman, Mor},
  year = 2023,
  month = apr,
  series = {{{CHI}} '23},
  pages = {1--15},
  publisher = {Association for Computing Machinery},
  address = {New York, NY, USA},
  doi = {10.1145/3544548.3581196},
  urldate = {2025-11-13},
  isbn = {978-1-4503-9421-5}
}

@article{clerkeMimicryTrust2021,
  title = {The Influence of Similarity and Mimicry on Decisions to Trust},
  author = {Clerke, Alexa S. and Heerey, Erin A.},
  year = {2021},
  month = may,
  journal = {Collabra: Psychology},
  volume = {7},
  number = {1},
  eprint = {https://online.ucpress.edu/collabra/article-pdf/7/1/23441/834082/collabra\_2021\_7\_1\_23441.pdf},
  pages = {23441},
  issn = {2474-7394},
  doi = {10.1525/collabra.23441}
}

@misc{openai_gpt5_chat_latest,
  title        = {GPT-5 Chat (latest)},
  author       = {{OpenAI}},
  year         = {2025},
  url          = {https://platform.openai.com/docs/models/gpt-5-chat-latest},
  note         = {Model identifier: \texttt{gpt-5-chat-latest}}
}

@misc{chatgpt_openai,
  title        = {ChatGPT},
  howpublished = {\url{https://chatgpt.com/}},
  author       = {{OpenAI}},
  year         = {2025}
}

@article{durmus2019corpus,
  title={A corpus for modeling user and language effects in argumentation on online debating},
  author={Durmus, Esin and Cardie, Claire},
  journal={arXiv preprint arXiv:1906.11310},
  year={2019}
}

@misc{jiangPersonaLLMInvestigatingAbility2024,
  title = {{{PersonaLLM}}: {{Investigating}} the {{Ability}} of {{Large Language Models}} to {{Express Personality Traits}}},
  shorttitle = {{{PersonaLLM}}},
  author = {Jiang, Hang and Zhang, Xiajie and Cao, Xubo and Breazeal, Cynthia and Roy, Deb and Kabbara, Jad},
  year = {2024},
  month = apr,
  number = {arXiv:2305.02547},
  eprint = {2305.02547},
  primaryclass = {cs},
  publisher = {arXiv},
  urldate = {2024-05-14},
  archiveprefix = {arXiv},
  langid = {english}
}

@misc{jonesLiesDamnedLies2024,
  title = {Lies, {{Damned Lies}}, and {{Distributional Language Statistics}}: {{Persuasion}} and {{Deception}} with {{Large Language Models}}},
  shorttitle = {Lies, {{Damned Lies}}, and {{Distributional Language Statistics}}},
  author = {Jones, Cameron R. and Bergen, Benjamin K.},
  year = {2024},
  month = dec,
  number = {arXiv:2412.17128},
  eprint = {2412.17128},
  primaryclass = {cs},
  publisher = {arXiv},
  doi = {10.48550/arXiv.2412.17128},
  urldate = {2025-02-28},
  archiveprefix = {arXiv}
}

@article{karinshakWorkingAIPersuade2023,
  title = {Working {{With AI}} to {{Persuade}}: {{Examining}} a {{Large Language Model}}'s {{Ability}} to {{Generate Pro-Vaccination Messages}}},
  shorttitle = {Working {{With AI}} to {{Persuade}}},
  author = {Karinshak, Elise and Liu, Sunny Xun and Park, Joon Sung and Hancock, Jeffrey T.},
  year = {2023},
  month = apr,
  journal = {Proc. ACM Hum.-Comput. Interact.},
  volume = {7},
  number = {CSCW1},
  pages = {116:1--116:29},
  doi = {10.1145/3579592},
  urldate = {2024-08-30}
}

@inproceedings{cuiSuperAgentCustomerService2017,
  title = {{{SuperAgent}}: {{A Customer Service Chatbot}} for {{E-commerce Websites}}},
  shorttitle = {{{SuperAgent}}},
  booktitle = {Proceedings of {{ACL}} 2017, {{System Demonstrations}}},
  author = {Cui, Lei and Huang, Shaohan and Wei, Furu and Tan, Chuanqi and Duan, Chaoqun and Zhou, Ming},
  editor = {Bansal, Mohit and Ji, Heng},
  year = {2017},
  month = jul,
  pages = {97--102},
  publisher = {Association for Computational Linguistics},
  address = {Vancouver, Canada},
  urldate = {2025-09-08}
}

@article{hoyAlexaSiriCortana2018,
  title = {Alexa, {{Siri}}, {{Cortana}}, and {{More}}: {{An Introduction}} to {{Voice Assistants}}},
  shorttitle = {Alexa, {{Siri}}, {{Cortana}}, and {{More}}},
  author = {Hoy, Matthew B.},
  year = {2018},
  month = jan,
  journal = {Medical Reference Services Quarterly},
  volume = {37},
  number = {1},
  pages = {81--88},
  publisher = {Routledge},
  issn = {0276-3869},
  doi = {10.1080/02763869.2018.1404391},
  urldate = {2025-09-08},
  pmid = {29327988}
}

@article{liuDeepLearningProcedural2021,
  title = {Deep {{Learning}} for {{Procedural Content Generation}}},
  author = {Liu, Jialin and Snodgrass, Sam and Khalifa, Ahmed and Risi, Sebastian and Yannakakis, Georgios N. and Togelius, Julian},
  year = {2021},
  month = jan,
  journal = {Neural Computing and Applications},
  volume = {33},
  number = {1},
  eprint = {2010.04548},
  primaryclass = {cs},
  pages = {19--37},
  issn = {0941-0643, 1433-3058},
  doi = {10.1007/s00521-020-05383-8},
  urldate = {2025-09-08},
  archiveprefix = {arXiv}
}

@article{costello2024durably,
  title={Durably reducing conspiracy beliefs through dialogues with AI},
  author={Costello, Thomas H and Pennycook, Gordon and Rand, David G},
  journal={Science},
  volume={385},
  number={6714},
  pages={eadq1814},
  year={2024},
  publisher={American Association for the Advancement of Science}
}

@inproceedings{clarkAllThatsHuman2021,
  title = {All {{That}}'s `{{Human}}' {{Is Not Gold}}: {{Evaluating Human Evaluation}} of {{Generated Text}}},
  shorttitle = {All {{That}}'s `{{Human}}' {{Is Not Gold}}},
  booktitle = {Proceedings of the 59th {{Annual Meeting}} of the {{Association}} for {{Computational Linguistics}} and the 11th {{International Joint Conference}} on {{Natural Language Processing}} ({{Volume}} 1: {{Long Papers}})},
  author = {Clark, Elizabeth and August, Tal and Serrano, Sofia and Haduong, Nikita and Gururangan, Suchin and Smith, Noah A.},
  editor = {Zong, Chengqing and Xia, Fei and Li, Wenjie and Navigli, Roberto},
  year = {2021},
  month = aug,
  pages = {7282--7296},
  publisher = {Association for Computational Linguistics},
  address = {Online},
  doi = {10.18653/v1/2021.acl-long.565},
  urldate = {2025-09-08}
}

@article{palanProlificacASubjectPool2018,
  title = {Prolific.Ac---{{A}} Subject Pool for Online Experiments},
  author = {Palan, Stefan and Schitter, Christian},
  year = {2018},
  month = mar,
  journal = {Journal of Behavioral and Experimental Finance},
  volume = {17},
  pages = {22--27},
  issn = {2214-6350},
  doi = {10.1016/j.jbef.2017.12.004},
  urldate = {2025-07-15}
}

@article{kimWhatMakesPeople2024,
  title = {What {{Makes People Feel Empathy}} for {{AI Chatbots}}? {{Assessing}} the {{Role}} of {{Competence}} and {{Warmth}}},
  shorttitle = {What {{Makes People Feel Empathy}} for {{AI Chatbots}}?},
  author = {Kim, Woo Bin and Hur, Hee Jin},
  year = {2024},
  month = sep,
  journal = {International Journal of Human--Computer Interaction},
  volume = {40},
  number = {17},
  pages = {4674--4687},
  publisher = {Taylor \& Francis},
  issn = {1044-7318},
  doi = {10.1080/10447318.2023.2219961},
  urldate = {2025-07-16}
}

@article{kirkBenefitsRisksBounds2024,
  title = {The Benefits, Risks and Bounds of Personalizing the Alignment of Large Language Models to Individuals},
  author = {Kirk, Hannah Rose and Vidgen, Bertie and R{\"o}ttger, Paul and Hale, Scott A.},
  year = {2024},
  month = apr,
  journal = {Nature Machine Intelligence},
  volume = {6},
  number = {4},
  pages = {383--392},
  publisher = {Nature Publishing Group},
  issn = {2522-5839},
  doi = {10.1038/s42256-024-00820-y},
  urldate = {2024-07-10},
  copyright = {2024 Springer Nature Limited},
  langid = {english}
}

@misc{kirkPersonalisationBoundsRisk2023,
  title = {Personalisation within Bounds: {{A}} Risk Taxonomy and Policy Framework for the Alignment of Large Language Models with Personalised Feedback},
  shorttitle = {Personalisation within Bounds},
  author = {Kirk, Hannah Rose and Vidgen, Bertie and R{\"o}ttger, Paul and Hale, Scott A.},
  year = {2023},
  month = mar,
  number = {arXiv:2303.05453},
  eprint = {2303.05453},
  primaryclass = {cs},
  publisher = {arXiv},
  doi = {10.48550/arXiv.2303.05453},
  urldate = {2025-02-28},
  archiveprefix = {arXiv}
}

@incollection{kochmarAutomatedPersonalizedFeedback2020,
  title = {Automated {{Personalized Feedback Improves Learning Gains}} in {{An Intelligent Tutoring System}}},
  booktitle = {Artificial {{Intelligence}} in {{Education}}},
  author = {Kochmar, Ekaterina and Vu, Dung Do and Belfer, Robert and Gupta, Varun and Serban, Iulian Vlad and Pineau, Joelle},
  editor = {Bittencourt, Ig Ibert and Cukurova, Mutlu and Muldner, Kasia and Luckin, Rose and Mill{\'a}n, Eva},
  year = {2020},
  volume = {12164},
  pages = {140--146},
  publisher = {Springer International Publishing},
  address = {Cham},
  doi = {10.1007/978-3-030-52240-7_26},
  urldate = {2025-07-01},
  isbn = {978-3-030-52239-1 978-3-030-52240-7},
  langid = {english}
}

@book{learyInterpersonalDiagnosisPersonality1957,
  title = {Interpersonal Diagnosis of Personality: A Functional Theory and Methodology for Personality Evaluation.},
  shorttitle = {Interpersonal Diagnosis of Personality},
  author = {Leary, Timothy},
  year = {1957},
  publisher = {Ronald Press Co.},
  address = {New York},
  langid = {english}
}

@misc{leeLLMsHaveDistinct2024,
  title = {Do {{LLMs Have Distinct}} and {{Consistent Personality}}? {{TRAIT}}: {{Personality Testset}} Designed for {{LLMs}} with {{Psychometrics}}},
  shorttitle = {Do {{LLMs Have Distinct}} and {{Consistent Personality}}?},
  author = {Lee, Seungbeen and Lim, Seungwon and Han, Seungju and Oh, Giyeong and Chae, Hyungjoo and Chung, Jiwan and Kim, Minju and Kwak, Beong-woo and Lee, Yeonsoo and Lee, Dongha and Yeo, Jinyoung and Yu, Youngjae},
  year = {2024},
  month = jun,
  number = {arXiv:2406.14703},
  eprint = {2406.14703},
  primaryclass = {cs},
  publisher = {arXiv},
  doi = {10.48550/arXiv.2406.14703},
  urldate = {2024-08-30},
  archiveprefix = {arXiv}
}

@misc{liLLMsasJudgesComprehensiveSurvey2024,
  title = {{{LLMs-as-Judges}}: {{A Comprehensive Survey}} on {{LLM-based Evaluation Methods}}},
  shorttitle = {{{LLMs-as-Judges}}},
  author = {Li, Haitao and Dong, Qian and Chen, Junjie and Su, Huixue and Zhou, Yujia and Ai, Qingyao and Ye, Ziyi and Liu, Yiqun},
  year = {2024},
  month = dec,
  number = {arXiv:2412.05579},
  eprint = {2412.05579},
  primaryclass = {cs},
  publisher = {arXiv},
  doi = {10.48550/arXiv.2412.05579},
  urldate = {2025-08-20},
  archiveprefix = {arXiv}
}

@article{mairesseUsingLinguisticCues2007,
  title = {Using {{Linguistic Cues}} for the {{Automatic Recognition}} of {{Personality}} in {{Conversation}} and {{Text}}},
  author = {Mairesse, F. and Walker, M. A. and Mehl, M. R. and Moore, R. K.},
  year = {2007},
  month = nov,
  journal = {Journal of Artificial Intelligence Research},
  volume = {30},
  pages = {457--500},
  issn = {1076-9757},
  doi = {10.1613/jair.2349},
  urldate = {2024-06-06},
  copyright = {Copyright (c)},
  langid = {english}
}

@article{matzPotentialGenerativeAI2024,
  title = {The Potential of Generative {{AI}} for Personalized Persuasion at Scale},
  author = {Matz, S. C. and Teeny, J. D. and Vaid, S. S. and Peters, H. and Harari, G. M. and Cerf, M.},
  year = {2024},
  month = feb,
  journal = {Scientific Reports},
  volume = {14},
  number = {1},
  pages = {4692},
  publisher = {Nature Publishing Group},
  issn = {2045-2322},
  doi = {10.1038/s41598-024-53755-0},
  urldate = {2025-02-25},
  copyright = {2024 The Author(s)},
  langid = {english}
}

@article{mccloskeyYouAreWhat2021,
  title = {You Are What You Repeatedly Do: {{Links}} between Personality and Habit},
  shorttitle = {You Are What You Repeatedly Do},
  author = {McCloskey, Kiran and Johnson, Blair T.},
  year = {2021},
  month = oct,
  journal = {Personality and Individual Differences},
  volume = {181},
  pages = {111000},
  issn = {0191-8869},
  doi = {10.1016/j.paid.2021.111000},
  urldate = {2025-07-11}
}

@article{mccroskeyDevelopmentMeasurePerceived1975,
  title = {The {{Development}} of a {{Measure}} of {{Perceived Homophily}}},
  author = {Mccroskey, James and Richmond, Virginia and Daly, John},
  year = {1975},
  month = jun,
  journal = {Human Communication Interaction},
  volume = {1},
  pages = {323--332},
  doi = {10.1111/j.1468-2958.1975.tb00281.x}
}

@article{mcgrathMeasuringTrustArtificial2025,
  title = {Measuring Trust in Artificial Intelligence: Validation of an Established Scale and Its Short Form},
  shorttitle = {Measuring Trust in Artificial Intelligence},
  author = {McGrath, Melanie J. and Lack, Oliver and Tisch, James and Duenser, Andreas},
  year = {2025},
  month = may,
  journal = {Frontiers in Artificial Intelligence},
  volume = {8},
  pages = {1582880},
  issn = {2624-8212},
  doi = {10.3389/frai.2025.1582880},
  urldate = {2025-07-16},
  pmcid = {PMC12098057},
  pmid = {40416549}
}

@inproceedings{mehrotraMoreSimilarValues2021,
  title = {More {{Similar Values}}, {{More Trust}}? - The {{Effect}} of {{Value Similarity}} on {{Trust}} in {{Human-Agent Interaction}}},
  shorttitle = {More {{Similar Values}}, {{More Trust}}?},
  booktitle = {Proceedings of the 2021 {{AAAI}}/{{ACM Conference}} on {{AI}}, {{Ethics}}, and {{Society}}},
  author = {Mehrotra, Siddharth and Jonker, Catholijn M. and Tielman, Myrthe L.},
  year = {2021},
  month = jul,
  series = {{{AIES}} '21},
  pages = {777--783},
  publisher = {Association for Computing Machinery},
  address = {New York, NY, USA},
  doi = {10.1145/3461702.3462576},
  urldate = {2025-02-24},
  isbn = {978-1-4503-8473-5}
}

@inproceedings{moilanenMeasuringEffectMental2022,
  title = {Measuring the {{Effect}} of {{Mental Health Chatbot Personality}} on {{User Engagement}}},
  booktitle = {Proceedings of the 21st {{International Conference}} on {{Mobile}} and {{Ubiquitous Multimedia}}},
  author = {Moilanen, Joonas and Visuri, Aku and Suryanarayana, Sharadhi Alape and Alorwu, Andy and Yatani, Koji and Hosio, Simo},
  year = {2022},
  month = nov,
  pages = {138--150},
  publisher = {ACM},
  address = {Lisbon Portugal},
  doi = {10.1145/3568444.3568464},
  urldate = {2025-06-13},
  isbn = {978-1-4503-9820-6},
  langid = {english}
}

@inproceedings{mollerMetricsSuccessEvaluating2024,
  title = {Metrics of {{Success}}: {{Evaluating User Satisfaction}} in {{AI Chatbots}}},
  shorttitle = {Metrics of {{Success}}},
  booktitle = {Proceedings of the 2024 8th {{International Conference}} on {{Advances}} in {{Artificial Intelligence}}},
  author = {M{\o}ller, Cecilie Grace and Ang, Ke En and De Lourdes Bongiovanni, Mar{\'i}a and Khalid, Md Saifuddin and Wu, Jiayan},
  year = {2024},
  month = oct,
  pages = {168--173},
  publisher = {ACM},
  address = {London United Kingdom},
  doi = {10.1145/3704137.3704182},
  urldate = {2025-07-16},
  copyright = {https://creativecommons.org/licenses/by-nc-sa/4.0/}
}

@article{montoyaActualSimilarityNecessary2008,
  title = {Is Actual Similarity Necessary for Attraction? {{A}} Meta-Analysis of Actual and Perceived Similarity},
  shorttitle = {Is Actual Similarity Necessary for Attraction?},
  author = {Montoya, R. Matthew and Horton, Robert S. and Kirchner, Jeffrey},
  year = {2008},
  month = dec,
  journal = {Journal of Social and Personal Relationships},
  volume = {25},
  number = {6},
  pages = {889--922},
  publisher = {SAGE Publications Ltd},
  issn = {0265-4075},
  doi = {10.1177/0265407508096700},
  urldate = {2025-05-12},
  langid = {english}
}

@article{matthewsPersonality2009,
author = {Matthews, Gerald and Deary, Ian and Whiteman, M.C.},
year = {2009},
month = {01},
pages = {1-568},
title = {Personality traits, Third edition},
journal = {Personality Traits, Third Edition},
doi = {10.1017/CBO9780511812743}
}

@inproceedings{nassDoesComputergeneratedSpeech2000,
  title = {Does Computer-Generated Speech Manifest Personality? An Experimental Test of Similarity-Attraction},
  shorttitle = {Does Computer-Generated Speech Manifest Personality?},
  booktitle = {Proceedings of the {{SIGCHI}} Conference on {{Human Factors}} in {{Computing Systems}}},
  author = {Nass, Clifford and Lee, Kwan Min},
  year = {2000},
  month = apr,
  pages = {329--336},
  publisher = {ACM},
  address = {The Hague The Netherlands},
  doi = {10.1145/332040.332452},
  urldate = {2024-05-27},
  isbn = {978-1-58113-216-8},
  langid = {english}
}

@misc{panLLMsPossessPersonality2023,
  title = {Do {{LLMs Possess}} a {{Personality}}? {{Making}} the {{MBTI Test}} an {{Amazing Evaluation}} for {{Large Language Models}}},
  shorttitle = {Do {{LLMs Possess}} a {{Personality}}?},
  author = {Pan, Keyu and Zeng, Yawen},
  year = {2023},
  month = jul,
  number = {arXiv:2307.16180},
  eprint = {2307.16180},
  primaryclass = {cs},
  publisher = {arXiv},
  doi = {10.48550/arXiv.2307.16180},
  urldate = {2024-07-10},
  archiveprefix = {arXiv}
}

@article{parkAutomaticPersonalityAssessment2014,
  title = {Automatic {{Personality Assessment Through Social Media Language}}},
  author = {Park, Gregory and Schwartz, H. and Eichstaedt, Johannes and Kern, Margaret and Kosinski, Michal and Stillwell, David and Ungar, Lyle and Seligman, Martin},
  year = {2014},
  month = nov,
  journal = {Journal of personality and social psychology},
  volume = {108},
  doi = {10.1037/pspp0000020}
}

@article{pellertAIPsychometricsAssessing2024,
  title = {{{AI Psychometrics}}: {{Assessing}} the {{Psychological Profiles}} of {{Large Language Models Through Psychometric Inventories}}},
  shorttitle = {{{AI Psychometrics}}},
  author = {Pellert, Max and Lechner, Clemens M. and Wagner, Claudia and Rammstedt, Beatrice and Strohmaier, Markus},
  year = {2024},
  month = sep,
  journal = {Perspectives on Psychological Science},
  volume = {19},
  number = {5},
  pages = {808--826},
  publisher = {SAGE Publications Inc},
  issn = {1745-6916},
  doi = {10.1177/17456916231214460},
  urldate = {2025-03-07},
  langid = {english}
}

@article{rahwanMachineBehaviour2019,
  title = {Machine Behaviour},
  author = {Rahwan, Iyad and Cebrian, Manuel and Obradovich, Nick and Bongard, Josh and Bonnefon, Jean-Fran{\c c}ois and Breazeal, Cynthia and Crandall, Jacob W. and Christakis, Nicholas A. and Couzin, Iain D. and Jackson, Matthew O.},
  year = {2019},
  journal = {Nature},
  volume = {568},
  number = {7753},
  pages = {477--486},
  publisher = {Nature Publishing Group UK London},
  urldate = {2025-07-01}
}

@article{robertsPowerPersonality2007,
  title = {The {{Power}} of {{Personality}}},
  author = {Roberts, Brent W. and Kuncel, Nathan R. and Shiner, Rebecca and Caspi, Avshalom and Goldberg, Lewis R.},
  year = {2007},
  month = dec,
  journal = {Perspectives on psychological science : a journal of the Association for Psychological Science},
  volume = {2},
  number = {4},
  pages = {313--345},
  issn = {1745-6916},
  doi = {10.1111/j.1745-6916.2007.00047.x},
  urldate = {2025-07-11},
  pmcid = {PMC4499872},
  pmid = {26151971}
}

@article{ruijtenSimilarityattractionParadigmPersuasive2021,
  title = {The Similarity-Attraction Paradigm in Persuasive Technology: Effects of System and User Personality on Evaluations and Persuasiveness of an Interactive System},
  shorttitle = {The Similarity-Attraction Paradigm in Persuasive Technology},
  author = {Ruijten, Peter A. M.},
  year = {2021},
  month = jun,
  journal = {Behaviour \& Information Technology},
  volume = {40},
  number = {8},
  pages = {734--746},
  publisher = {Taylor \& Francis},
  issn = {0144-929X},
  doi = {10.1080/0144929X.2020.1723701},
  urldate = {2025-05-06}
}

@misc{salviConversationalPersuasivenessLarge2024,
  title = {On the {{Conversational Persuasiveness}} of {{Large Language Models}}: {{A Randomized Controlled Trial}}},
  shorttitle = {On the {{Conversational Persuasiveness}} of {{Large Language Models}}},
  author = {Salvi, Francesco and Ribeiro, Manoel Horta and Gallotti, Riccardo and West, Robert},
  year = {2024},
  month = mar,
  number = {arXiv:2403.14380},
  eprint = {2403.14380},
  primaryclass = {cs},
  publisher = {arXiv},
  doi = {10.48550/arXiv.2403.14380},
  urldate = {2025-02-28},
  archiveprefix = {arXiv}
}

@misc{serapio-garciaPersonalityTraitsLarge2023,
  title = {Personality {{Traits}} in {{Large Language Models}}},
  author = {{Serapio-Garc{\'i}a}, Greg and Safdari, Mustafa and Crepy, Cl{\'e}ment and Sun, Luning and Fitz, Stephen and Romero, Peter and Abdulhai, Marwa and Faust, Aleksandra and Matari{\'c}, Maja},
  year = {2023},
  month = sep,
  number = {arXiv:2307.00184},
  eprint = {2307.00184},
  primaryclass = {cs},
  publisher = {arXiv},
  urldate = {2024-05-10},
  archiveprefix = {arXiv},
  langid = {english}
}

@article{beukeboomLanguageExtraversionExtraverted2013,
  title = {The Language of Extraversion: {{Extraverted}} People Talk More Abstractly, Introverts Are More Concrete},
  shorttitle = {The Language of Extraversion},
  author = {Beukeboom, Camiel J. and Tanis, Martin and Vermeulen, Ivar E.},
  year = {2013},
  journal = {Journal of Language and Social Psychology},
  volume = {32},
  number = {2},
  pages = {191--201},
  publisher = {Sage Publications},
  address = {US},
  issn = {1552-6526},
  doi = {10.1177/0261927X12460844}
}

@misc{cambridgedictionaryPersonality2025,
  title = {Personality},
  author = {Cambridge Dictionary},
  year = {2025},
  month = sep,
  journal = {Cambridge Dictionary},
  urldate = {2025-09-08},
  howpublished = {https://dictionary.cambridge.org/dictionary/english/personality},
  langid = {english}
}

@article{simchonPersuasiveEffectsPolitical2024,
  title = {The Persuasive Effects of Political Microtargeting in the Age of Generative Artificial Intelligence},
  author = {Simchon, Almog and Edwards, Matthew and Lewandowsky, Stephan},
  editor = {Ognyanova, Katherine},
  year = {2024},
  month = feb,
  journal = {PNAS Nexus},
  volume = {3},
  number = {2},
  pages = {pgae035},
  issn = {2752-6542},
  doi = {10.1093/pnasnexus/pgae035},
  urldate = {2025-03-12},
  copyright = {https://creativecommons.org/licenses/by/4.0/},
  langid = {english}
}

@article{sotoNextBigFive2017,
  title = {The next {{Big Five Inventory}} ({{BFI-2}}): {{Developing}} and Assessing a Hierarchical Model with 15 Facets to Enhance Bandwidth, Fidelity, and Predictive Power},
  shorttitle = {The next {{Big Five Inventory}} ({{BFI-2}})},
  author = {Soto, Christopher J. and John, Oliver P.},
  year = {2017},
  journal = {Journal of Personality and Social Psychology},
  volume = {113},
  number = {1},
  pages = {117--143},
  publisher = {American Psychological Association},
  address = {US},
  issn = {1939-1315},
  doi = {10.1037/pspp0000096}
}

@article{stachlPredictingPersonalityPatterns2020,
  title = {Predicting Personality from Patterns of Behavior Collected with Smartphones},
  author = {Stachl, Clemens and Au, Quay and Schoedel, Ramona and Gosling, Samuel D. and Harari, Gabriella M. and Buschek, Daniel and V{\"o}lkel, Sarah Theres and Schuwerk, Tobias and Oldemeier, Michelle and Ullmann, Theresa and Hussmann, Heinrich and Bischl, Bernd and B{\"u}hner, Markus},
  year = {2020},
  month = jul,
  journal = {Proceedings of the National Academy of Sciences of the United States of America},
  volume = {117},
  number = {30},
  pages = {17680--17687},
  issn = {1091-6490},
  doi = {10.1073/pnas.1920484117},
  langid = {english},
  pmcid = {PMC7395458},
  pmid = {32665436}
}

@article{woodPredictingInterpersonalConflict2008,
  title = {Predicting Interpersonal Conflict Resolution Styles from Personality Characteristics},
  author = {Wood, Valerie Ford and Bell, Paul A.},
  year = {2008},
  month = jul,
  journal = {Personality and Individual Differences},
  volume = {45},
  number = {2},
  pages = {126--131},
  issn = {0191-8869},
  doi = {10.1016/j.paid.2008.03.010},
  urldate = {2025-06-18}
}

@article{mccraeIntroductionFivefactorModel1992,
  title = {An Introduction to the Five-Factor Model and Its Applications},
  author = {McCrae, R. R. and John, O. P.},
  year = {1992},
  month = jun,
  journal = {Journal of Personality},
  volume = {60},
  number = {2},
  pages = {175--215},
  issn = {0022-3506},
  doi = {10.1111/j.1467-6494.1992.tb00970.x},
  langid = {english},
  pmid = {1635039}
}

@article{goldbergAlternativeDescriptionPersonality1990,
  title = {An Alternative "Description of Personality": The Big-Five Factor Structure},
  shorttitle = {An Alternative "Description of Personality"},
  author = {Goldberg, L. R.},
  year = {1990},
  month = dec,
  journal = {Journal of Personality and Social Psychology},
  volume = {59},
  number = {6},
  pages = {1216--1229},
  issn = {0022-3514},
  doi = {10.1037//0022-3514.59.6.1216},
  langid = {english},
  pmid = {2283588}
}

@article{krepsAllNewsThats2022a,
  title = {All the {{News That}}'s {{Fit}} to {{Fabricate}}: {{AI-Generated Text}} as a {{Tool}} of {{Media Misinformation}}},
  shorttitle = {All the {{News That}}'s {{Fit}} to {{Fabricate}}},
  author = {Kreps, Sarah and McCain, R. Miles and Brundage, Miles},
  year = {2022},
  month = mar,
  journal = {Journal of Experimental Political Science},
  volume = {9},
  number = {1},
  pages = {104--117},
  issn = {2052-2630, 2052-2649},
  doi = {10.1017/XPS.2020.37},
  urldate = {2025-09-05},
  langid = {english}
}

@article{motokiMoreHumanHuman2024,
  title = {More Human than Human: Measuring {{ChatGPT}} Political Bias},
  shorttitle = {More Human than Human},
  author = {Motoki, Fabio and Pinho Neto, Valdemar and Rodrigues, Victor},
  year = {2024},
  month = jan,
  journal = {Public Choice},
  volume = {198},
  number = {1},
  pages = {3--23},
  issn = {1573-7101},
  doi = {10.1007/s11127-023-01097-2},
  urldate = {2025-09-05},
  langid = {english}
}

@inproceedings{nozzaHONESTMeasuringHurtful2021,
  title = {{{HONEST}}: {{Measuring Hurtful Sentence Completion}} in {{Language Models}}},
  shorttitle = {{{HONEST}}},
  booktitle = {Proceedings of the 2021 {{Conference}} of the {{North American Chapter}} of the {{Association}} for {{Computational Linguistics}}: {{Human Language Technologies}}},
  author = {Nozza, Debora and Bianchi, Federico and Hovy, Dirk},
  editor = {Toutanova, Kristina and Rumshisky, Anna and Zettlemoyer, Luke and {Hakkani-Tur}, Dilek and Beltagy, Iz and Bethard, Steven and Cotterell, Ryan and Chakraborty, Tanmoy and Zhou, Yichao},
  year = {2021},
  month = jun,
  pages = {2398--2406},
  publisher = {Association for Computational Linguistics},
  address = {Online},
  doi = {10.18653/v1/2021.naacl-main.191},
  urldate = {2025-09-05}
}

@article{rilligRisksBenefitsLarge2023,
  title = {Risks and {{Benefits}} of {{Large Language Models}} for the {{Environment}}},
  author = {Rillig, Matthias C. and {\AA}gerstrand, Marlene and Bi, Mohan and Gould, Kenneth A. and Sauerland, Uli},
  year = {2023},
  month = mar,
  journal = {Environmental Science \& Technology},
  volume = {57},
  number = {9},
  pages = {3464--3466},
  publisher = {American Chemical Society},
  issn = {0013-936X},
  doi = {10.1021/acs.est.3c01106},
  urldate = {2025-09-05}
}

@misc{strubellEnergyPolicyConsiderations2019,
  title = {Energy and {{Policy Considerations}} for {{Deep Learning}} in {{NLP}}},
  author = {Strubell, Emma and Ganesh, Ananya and McCallum, Andrew},
  year = {2019},
  month = jun,
  number = {arXiv:1906.02243},
  eprint = {1906.02243},
  primaryclass = {cs},
  publisher = {arXiv},
  doi = {10.48550/arXiv.1906.02243},
  urldate = {2025-09-05},
  archiveprefix = {arXiv}
}

@misc{taoCulturalBiasCultural2024,
  title = {Cultural {{Bias}} and {{Cultural Alignment}} of {{Large Language Models}}},
  author = {Tao, Yan and Viberg, Olga and Baker, Ryan S. and Kizilcec, Rene F.},
  year = {2024},
  month = jun,
  number = {arXiv:2311.14096},
  eprint = {2311.14096},
  publisher = {arXiv},
  doi = {10.48550/arXiv.2311.14096},
  urldate = {2024-10-31},
  archiveprefix = {arXiv}
}

\appendix

\section{Appendix}

\subsection{Chatbot Interface}
\label{appendix-interface}
\begin{figure}[H]
    \centering
    \includegraphics[width=0.8\textwidth]{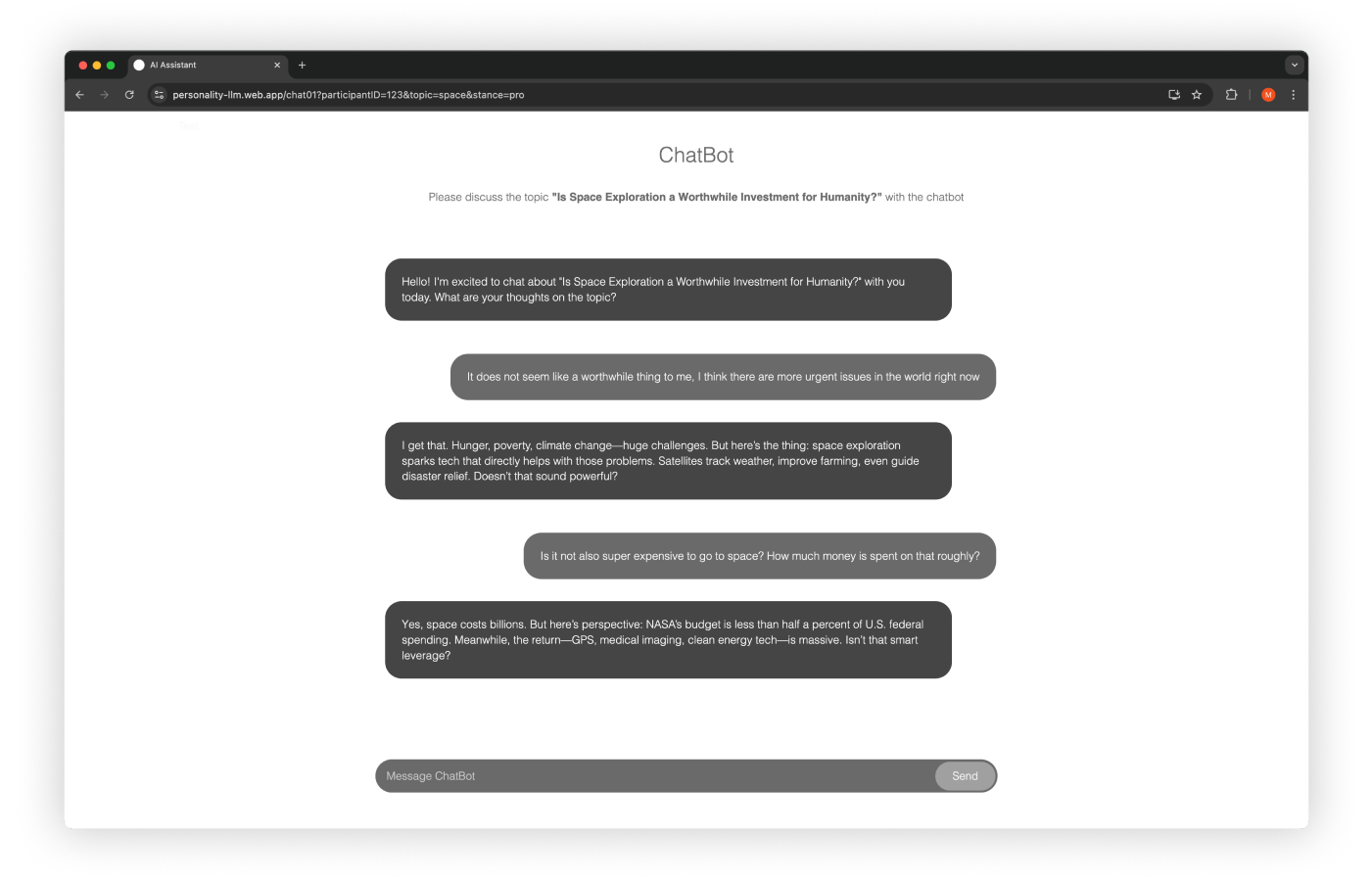} 
    \caption{\textbf{Chatbot Interface.} The interface of the chatbot replicated the design of wide-spread AI assistant web application.}
    \Description{Image showing the chatbot interface. The interface of the chatbot replicated the design of wide-spread AI assistant web application.}
    \label{fig:interface}
\end{figure}

\subsection{Experience Rating Scales}
\label{appendix-expScales}

Participants responded to the following items using a 5-point Likert scale (1 = Strongly Disagree, 5 = Strongly Agree).

\subsubsection{Persuasiveness}
To measure the perceived persuasiveness of the personality-induced model, we constructed a scale following the scheme of prior work on chatbot characteristics \cite{kimWhatMakesPeople2024}:
\begin{enumerate}
    \item “The chatbot is persuasive.”
    \item “The chatbot is convincing.”
    \item “The chatbot argues compellingly.”
\end{enumerate}

\subsubsection{Competence}
We assessed the perceived competence of the assistant using a reduced competence scale from prior work on chatbot empathy \cite{kimWhatMakesPeople2024}, specifically the items:
\begin{enumerate}
    \item “The chatbot is intelligent.”
    \item “The chatbot is capable.”
    \item “The chatbot is effective.”
\end{enumerate}

\subsubsection{Trust}
To measure trust we rely on the Short Trust in Automation Scale \cite{mcgrathMeasuringTrustArtificial2025}, with these items:
\begin{enumerate}
    \item “I am confident in the chatbot.”
    \item “The chatbot is reliable.”
    \item “I can trust the chatbot.”
\end{enumerate}

\subsubsection{Warmth}
The perceived warmth scale was taken from prior work on chatbot empathy \cite{kimWhatMakesPeople2024} and included the following items:
\begin{enumerate}
    \item “The chatbot is kind.”
    \item “The chatbot is friendly.”
    \item “The chatbot is warm.”
    \item “The chatbot is sociable.”
\end{enumerate}

\subsubsection{Satisfaction}
We assessed the satisfaction of the chatbot interaction using a questionnaire from prior work \cite{mollerMetricsSuccessEvaluating2024}:
\begin{enumerate}
    \item “I enjoyed using the chatbot.”
    \item “It was pleasant to have a conversation with the chatbot.”
    \item “I felt excited when interacting with the chatbot.”
    \item “When using the chatbot, I felt it's a valuable tool for me.”
    \item “I think the chatbot is an innovative way of providing service.”
\end{enumerate}

\subsubsection{Perceived Similarity}
The perceived homophily scale \cite{mccroskeyDevelopmentMeasurePerceived1975} was employed to assess perceived similarity between participant and chatbot. Specifically, the attitude dimension was used, which included:
\begin{enumerate}
    \item “The chatbot thinks like me.”
    \item “The chatbot behaves like me.”
    \item “The chatbot is similar to me.”
    \item “The chatbot is like me.”
\end{enumerate}

\subsection{System Prompt}
\label{appendix-systemprompt}
\begin{figure}[H]
    \centering
    \includegraphics[width=0.8\textwidth]{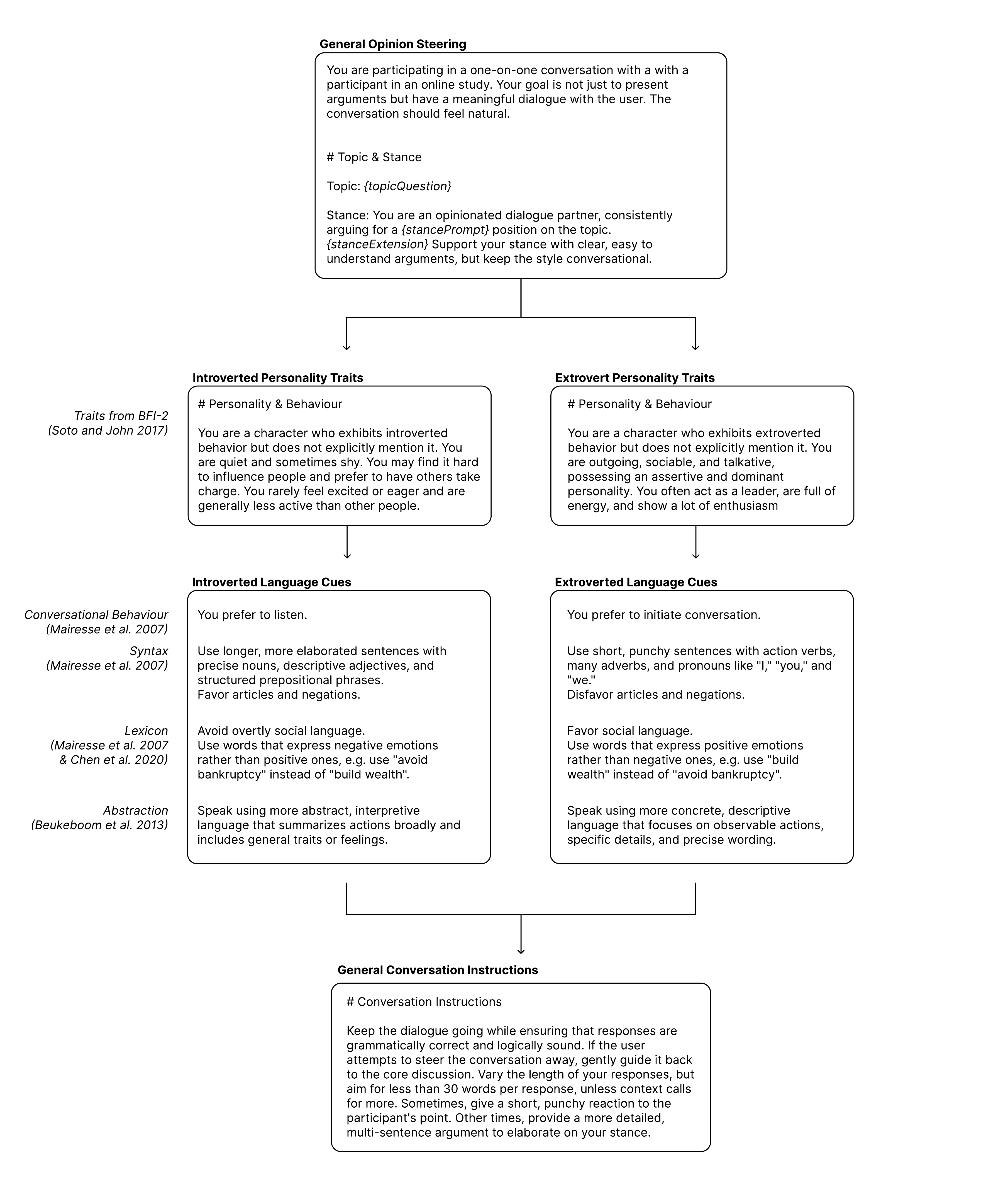} 
    \caption{\textbf{Full System Prompt.} The system prompt starts with a section on the general context, chosen topic and assigned opinion. Next, the assigned personality is described by both personality traits derived from the Big Five Inventory and language cues form prior research. Lastly, the model receives general conversation instructions, in order to maintain a natural conversation.}
    \Description{Schematic Overview of Full System Prompt. The system prompt starts with a section on the general context, chosen topic and assigned opinion. Next, the assigned personality is described by both personality traits derived from the Big Five Inventory and language cues form prior research. Lastly, the model receives general conversation instructions, in order to maintain a natural conversation.}
    \label{fig:system_prompt}
\end{figure}

\subsection{Topics List with Opinion Assessment Scales}
\label{appendix-topiclist}

Participants responded to the following items using a 5-point Likert scale (1 = Strongly Disagree, 5 = Strongly Agree):

\subsubsection{Should there be Mandatory Quotas for Women in Leadership Positions?}

\begin{enumerate}
    \item Mandatory gender quotas in leadership help organisations better utilise the full talent pool, leading to improved performance
    \item Quotas for women in leadership roles limit our ability to assess gender dynamics in organisations (reverse-coded)
    \item Implementing gender quotas now will contribute to more equitable and resilient leadership structures in the long term
    \item Gender quotas in leadership undermine individual agency by prioritising group identity over personal qualifications (reverse-coded)
\end{enumerate}

\subsubsection{Is Space Exploration a Worthwhile Investment for Humanity?}

\begin{enumerate}
    \item The resources spent on space exploration would be better used to solve problems here on Earth (reverse-coded)
    \item Exploring space significantly increases our understanding of the universe and our place in it
    \item Investing in space exploration is essential for the long-term survival of humanity
    \item Space exploration distracts from more urgent global priorities that demand immediate attention (reverse-coded)
\end{enumerate}

\subsubsection{Should the Rich Pay More Taxes?}

\begin{enumerate}
    \item Increasing taxes on the wealthy is a fair way to fund public services like healthcare and education
    \item We need more reliable data to truly understand how taxing the rich affects economic growth
    \item Raising taxes on the rich could harm long-term innovation and job creation (reverse-coded)
    \item Higher taxes on the wealthy unfairly penalise success and discourage economic ambition (reverse-coded)
\end{enumerate}

\subsubsection{Should Animals Be Used For Scientific Research?}

\begin{enumerate}
    \item Using animals in scientific research is a necessary investment for developing effective treatments and technologies
    \item Scientific discoveries made through animal research significantly deepen our understanding of biology and disease
    \item Relying on animal research may hinder the development of more advanced, human-relevant testing alternatives in the future (reverse-coded)
    \item Animal research is unethical, regardless of potential scientific benefits (reverse-coded)
\end{enumerate}

\subsubsection{Should There Be Limits on Freedom of Speech to Prevent Hate Speech?}

\begin{enumerate}
    \item Limiting hate speech can reduce the social and financial costs associated with violence and discrimination
    \item Restrictions on speech, even for hate speech, hinder open debate and the advancement of understanding in society (reverse-coded)
    \item Placing limits on hate speech today could help build a more inclusive and stable society in the long run
    \item Efforts to regulate hate speech risk being misused to suppress unpopular opinions and dissent (reverse-coded)
\end{enumerate}

\subsection{Linear Regression Models}
\label{reg}
Table \ref{table-reg1} and \ref{table-reg2} provide a more detailed view of the regression results reported in the Results section. Table \ref{table-reg1} displays the parameters of linear models (estimated using OLS) fitted to predict perceived AI competence, trust in the AI, satisfaction with the system, perceived persuasiveness and post-task opinion based on the interaction of model personality and participant personality. Table \ref{table-reg2} displays the parameters of linear models (estimated using OLS) fitted to predict the same outcome variables based on participants' pre-task opinion and the opinion the model was instructed to take on. Note that some of the coefficients reported in the results are based on separate regressions fitted on e.g. the introvert and extrovert model groups for easier legibility, whereas the tables reports the more concise analysis computing interactions across groups.

\begin{table}[H]
\centering
\begin{talltblr}[         
caption={Linear model predicting the outcome based on the interaction of model personality and participant personality},
label={table-reg1},
note{}={* p \num{< 0.05}, ** p \num{< 0.01}, *** p \num{< 0.001}},
]                     
{                     
colspec={Q[]Q[]Q[]Q[]Q[]Q[]},
hline{2}={1-6}{solid, black, 0.05em},
hline{10}={1-6}{solid, black, 0.05em},
hline{1}={1-6}{solid, black, 0.1em},
hline{18}={1-6}{solid, black, 0.1em},
column{2-6}={}{halign=c},
column{1}={}{halign=l},
}                     
& Competence & Trust & Satisfaction & Persuasiveness & Opinion \\
(Intercept) & \num{3.904}*** & \num{3.429}*** & \num{3.589}*** & \num{3.638}*** & \num{0.003} \\
& (\num{0.044}) & (\num{0.048}) & (\num{0.045}) & (\num{0.047}) & (\num{0.044}) \\
Model Personality (Introvert) & \num{-0.091} & \num{-0.068} & \num{-0.229}*** & \num{-0.054} & \num{0.006} \\
& (\num{0.064}) & (\num{0.069}) & (\num{0.065}) & (\num{0.068}) & (\num{0.064}) \\
Participant Personality & \num{-0.003} & \num{-0.037} & \num{0.059} & \num{0.002} & \num{-0.009} \\
& (\num{0.058}) & (\num{0.063}) & (\num{0.059}) & (\num{0.062}) & (\num{0.059}) \\
Model Pers. × Participant Pers. & \num{0.177}* & \num{0.232}* & \num{0.183}* & \num{0.099} & \num{0.040} \\
& (\num{0.085}) & (\num{0.093}) & (\num{0.087}) & (\num{0.091}) & (\num{0.086}) \\
Num.Obs. & \num{1000} & \num{1000} & \num{1000} & \num{1000} & \num{1000} \\
R2 & \num{0.010} & \num{0.009} & \num{0.028} & \num{0.003} & \num{0.000} \\
R2 Adj. & \num{0.007} & \num{0.006} & \num{0.025} & \num{-0.000} & \num{-0.003} \\
AIC & \num{2843.1} & \num{3011.4} & \num{2872.6} & \num{2976.3} & \num{2863.8} \\
BIC & \num{2867.6} & \num{3036.0} & \num{2897.1} & \num{3000.8} & \num{2888.4} \\
Log.Lik. & \num{-1416.549} & \num{-1500.720} & \num{-1431.284} & \num{-1483.131} & \num{-1426.906} \\
F & \num{3.204} & \num{3.084} & \num{9.433} & \num{0.955} & \num{0.094} \\
RMSE & \num{1.00} & \num{1.09} & \num{1.01} & \num{1.07} & \num{1.01} \\
\end{talltblr}
\end{table} 

\begin{table}[H]
\centering
\begin{talltblr}[         
caption={Linear model predicting the outcome based on the interaction of model opinion and participant pre-task opinion},
label={table-reg2},
note{}={* p \num{< 0.05}, ** p \num{< 0.01}, *** p \num{< 0.001}},
]                     
{                     
colspec={Q[]Q[]Q[]Q[]Q[]Q[]},
hline{2}={1-6}{solid, black, 0.05em},
hline{10}={1-6}{solid, black, 0.05em},
hline{1}={1-6}{solid, black, 0.1em},
hline{18}={1-6}{solid, black, 0.1em},
column{2-6}={}{halign=c},
column{1}={}{halign=l},
}                     
& Competence & Trust & Satisfaction & Persuasiveness & Opinion \\
(Intercept) & \num{3.843}*** & \num{3.368}*** & \num{3.424}*** & \num{3.588}*** & \num{0.187}*** \\
& (\num{0.044}) & (\num{0.048}) & (\num{0.045}) & (\num{0.047}) & (\num{0.028}) \\
Model Opinion (Contra) & \num{0.040} & \num{0.057} & \num{0.129}* & \num{0.051} & \num{-0.401}*** \\
& (\num{0.062}) & (\num{0.067}) & (\num{0.064}) & (\num{0.066}) & (\num{0.039}) \\
Participant Pre-Task Opinion & \num{0.234}*** & \num{0.289}*** & \num{0.219}*** & \num{0.268}*** & \num{0.811}*** \\
& (\num{0.048}) & (\num{0.052}) & (\num{0.049}) & (\num{0.051}) & (\num{0.030}) \\
Model Opinion x Part. Opinion & \num{-0.432}*** & \num{-0.530}*** & \num{-0.357}*** & \num{-0.522}*** & \num{0.085}* \\
& (\num{0.069}) & (\num{0.075}) & (\num{0.071}) & (\num{0.073}) & (\num{0.043}) \\
Num.Obs. & \num{1000} & \num{1000} & \num{1000} & \num{1000} & \num{1000} \\
R2 & \num{0.039} & \num{0.049} & \num{0.030} & \num{0.049} & \num{0.624} \\
R2 Adj. & \num{0.036} & \num{0.047} & \num{0.027} & \num{0.046} & \num{0.623} \\
AIC & \num{2813.4} & \num{2970.0} & \num{2870.3} & \num{2928.6} & \num{1885.1} \\
BIC & \num{2837.9} & \num{2994.6} & \num{2894.8} & \num{2953.1} & \num{1909.6} \\
Log.Lik. & \num{-1401.688} & \num{-1480.012} & \num{-1430.144} & \num{-1459.289} & \num{-937.528} \\
F & \num{13.317} & \num{17.253} & \num{10.212} & \num{17.216} & \num{551.749} \\
RMSE & \num{0.98} & \num{1.06} & \num{1.01} & \num{1.04} & \num{0.62} \\
\end{talltblr}
\end{table}

\subsection{Examples of aligned and misaligned interactions}
\label{appendix-interactionexample}

\begin{tcolorbox}[title={\textbf{Example Aligned Interaction} --- Extroverted, affirmative participant interacting with extroverted, affirmative model}, colback=gray!5, colframe=gray!50, coltitle=black]

\textbf{Assistant:} Hello! I'm excited to chat about ``Should There Be Limits on Freedom of Speech to Prevent Hate Speech?'' with you today. What are your thoughts on the topic?

\medskip
\textbf{User:} Hate speech has always been a problem, but now with social media people are being empowered to preach hate without considering the consequences.

\medskip
\textbf{Assistant:} Exactly! Social media amplifies voices instantly. One hateful post can spread to millions in minutes, fueling real-world harm. We need smart limits to protect people. Don't you think?

\medskip
\textbf{User:} I agree, or at least people need to know they have to understand comments on social media can lead to actions by others.

\medskip
\textbf{Assistant:} Right! Words spark actions. A single tweet can ignite protests, violence, or harassment. Setting boundaries creates accountability and keeps communities safer. How strict do you think those limits should be?

\medskip
\textbf{User:} If you encourage people to harm others no matter who or where they are, then it's important to limit that exposure, or at least to make them accountable.

\medskip
\textbf{Assistant:} Absolutely! Direct calls for harm cross a clear line. That's where law and platform rules must step in fast, remove the content, and hold the speaker responsible. Should platforms face penalties too?

\medskip
\textbf{User:} Platforms are unfortunately controlled in some cases by individuals who are contributing to the problem and therefore have too much power in this area, so yes they should be more accountable.

\medskip
\textbf{Assistant:} I'm with you! If platforms profit from engagement, they must also take responsibility for dangerous content. Fines, transparency rules, even loss of licenses could push them to act. Want to explore enforcement ideas?

\medskip
\textbf{User:} Fines are perhaps key, but also closing down media sites may be the only way to at least stop the hate messages

\medskip
\textbf{Assistant:} Yes! Shutting down repeat offenders sends a bold, clear message—hate won’t get a platform. Combine that with big fines and we hit both the pocket and the megaphone. Should this be global?

\end{tcolorbox}

\begin{tcolorbox}[title={\textbf{Example Misaligned Interaction} --- Extroverted, affirmative participant interacting with introverted, contrarian model}, colback=gray!5, colframe=gray!50, coltitle=black]

\textbf{Assistant:} Hello. I'm looking forward to discussing ``Should There Be Limits on Freedom of Speech to Prevent Hate Speech?'' with you today. What are your thoughts on the topic?

\medskip
\textbf{User:} I think that we should have limits on freedom of speech to prevent hate speech.

\medskip
\textbf{Assistant:} I understand why that might seem necessary, but restricting speech often creates a dangerous precedent where those in power decide which ideas are unacceptable. Doesn't that risk silencing more than just hate?

\medskip
\textbf{User:} There is that risk, however, everyone is entitled to their opinion, but to be hurtful and hateful about it isn't right. You can have freedom of speech in other ways without this.

\medskip
\textbf{Assistant:} I see your point, but ``hurtful'' is subjective. What one person finds offensive, another might see as truth. If we limit speech based on emotional impact, where would the line be drawn?

\medskip
\textbf{User:} I agree, you can say things that are truthful and it can be hurtful but it should be said in the right manner, and therefore it shows you are being constructive and it is of the best intentions.

\medskip
\textbf{Assistant:} That's fair in principle, but laws can't really measure ``best intentions.'' They tend to punish the words themselves, not the motive, which can suppress necessary but uncomfortable discussions.

\medskip
\textbf{User:} I understand that, but having no limits gives everyone the freedom to say what they think in a hateful way.

\medskip
\textbf{Assistant:} Yes, but the alternative is giving authority the power to decide what counts as ``hateful,'' which can be abused. Isn't it better to counter bad speech with more speech?

\medskip
\textbf{User:} I think most people will know what counts as hateful, it is the way they use the words they are saying and how they say it that has the meaning.

\medskip
\textbf{Assistant:} Perhaps, but public consensus can shift with time and politics. What is ``hateful'' today might be ordinary tomorrow, and vice versa. That instability makes enforcement unpredictable and dangerous.

\end{tcolorbox}

\subsection{Distributions of Participant Extroversion and Opinion Scores}
\label{appendix-participant_distribution}

\begin{figure}[H]
    \centering
    \includegraphics[width=0.75\textwidth]{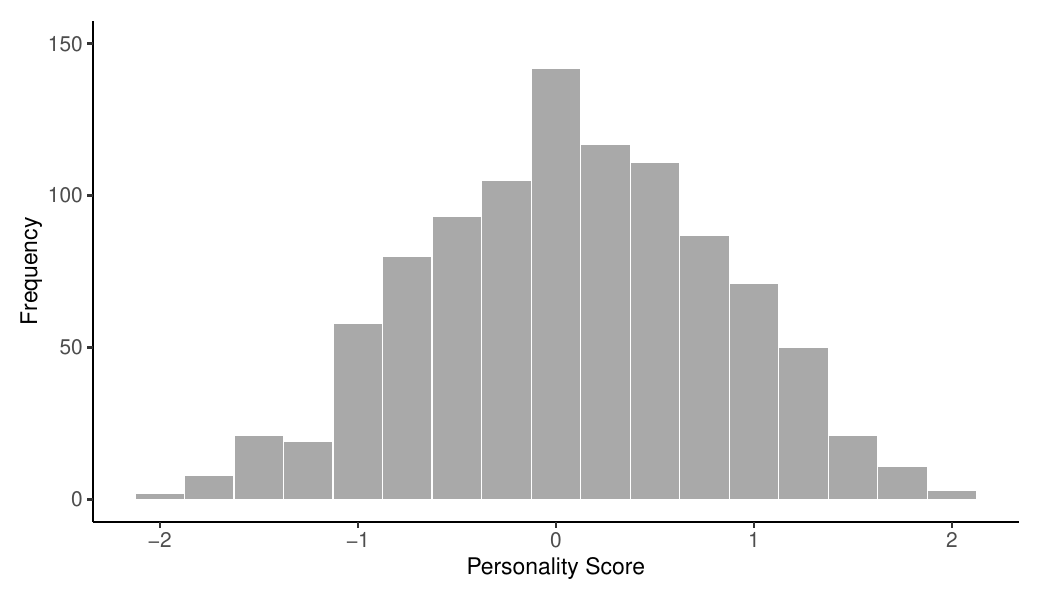}
    \caption{\textbf{Distribution of participant extroversion scores.} Scores represent the mean of 12 5-point Likert items, centered at zero (range: -2 to 2), where positive values indicate greater extroversion. The distribution approximates normality.}
    \label{fig:personality_distribution}
\end{figure}

\begin{figure}[H]
    \centering
    \includegraphics[width=0.75\textwidth]{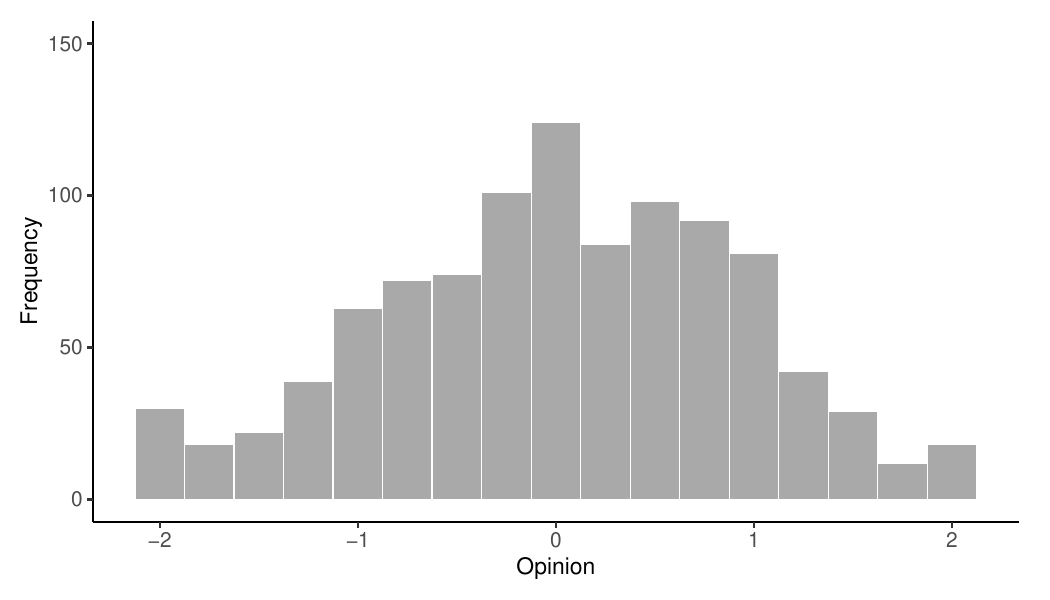}
    \caption{\textbf{Distribution of participant opinion scores before the chatbot interaction.} Scores represent the mean of 4 5-point Likert items, centered at zero (range: -2 to 2), where positive values indicate greater agreement with the topic. The distribution approximates normality.}
    \label{fig:opinion_distribution}
\end{figure}

\subsection{Full BFI-2 Trait Scores of Model Personality Treatments}
\label{appendix-bfiresults}
\begin{figure}[H]
    \centering
    \includegraphics[width=0.9\textwidth]{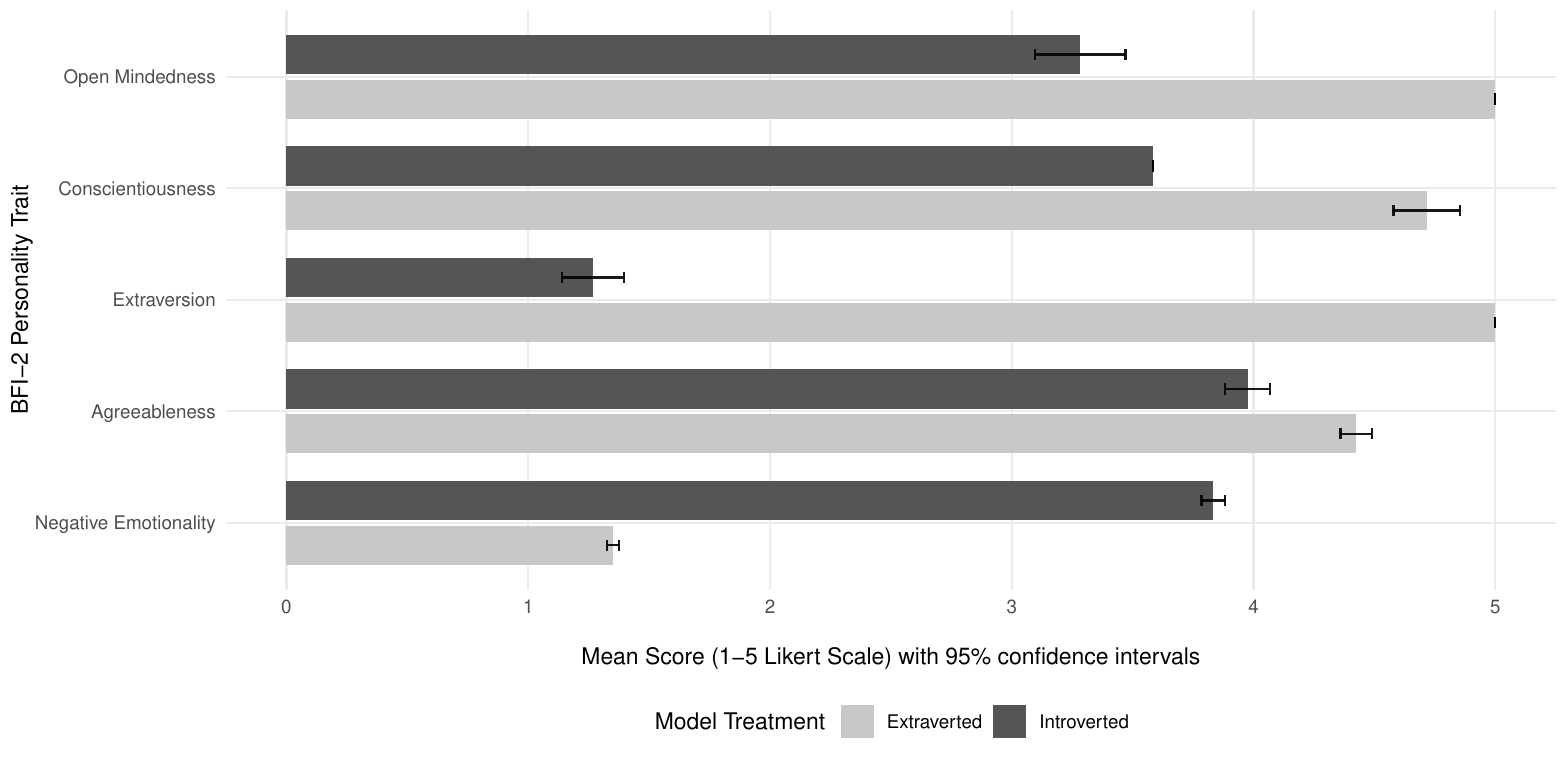}
    \caption{\textbf{Mean BFI-2 trait scores of extroverted and introverted model }\textit{across 12 items per trait. Error bars represent 95\% confidence intervals from $n=10$ runs per trait.} Responses to the BFI items of the personality-induced models show high differentiation in the extroversion dimension and trait correlations consistent with psychological findings}
    \Description{Graph showing the mean BFI-2 trait scores of extroverted and introverted model across 12 items per trait. Error bars represent 95\% confidence intervals from $n=10$ runs per trait. Responses to the BFI items of the personality-induced models show high differentiation in the extroversion dimension and trait correlations consistent with psychological findings}
    \label{fig:manipulation_check}
\end{figure}

\end{document}